\documentclass[11pt]{article}

\PassOptionsToPackage{nosort}{cite}

\usepackage{amssymb}
\usepackage{chet}
\usepackage{mathtools}
\usepackage[scale=0.86,defaultsans]{opensans}
\usepackage{booktabs}
\usepackage{tablefootnote}

\definecolor{darkblue}{rgb}{0.1,0.1,0.7}
\hypersetup{colorlinks,
           linkcolor={darkblue},
           citecolor={darkblue},
           urlcolor={darkblue}
}

\newcommand{\veps}{\ensuremath{\varepsilon}\xspace}
\newcommand{\cC}{\ensuremath{\mathcal{C}}\xspace}

\newcommand{\vv}{\ensuremath{\mathbf{v}}}
\newcommand{\lsp}{\hspace{1pt}}
\newcommand{\llsp}{\hspace{0.5pt}}

\newcommand{\llnsp}{\hspace{-0.5pt}}
\newcommand{\vvi}{\ensuremath{\mathbf{v}_{\hspace{-0.8pt}i}}}
\newcommand{\vvj}{\ensuremath{\mathbf{v}_{\hspace{-1.1pt}j}}}
\newcommand{\Bconst}{B}
\DeclareMathOperator{\tr}{tr}

\usepackage[normalem]{ulem}

\def\ge{\geqslant}
\def\le{\leqslant}

\newcommand{\reef}[1]{(\ref{#1})}
\def\vareps{\varepsilon}

\def\veps{\varepsilon}

\newcommand{\beq}{\begin{equation}}
\newcommand{\eeq}{\end{equation}}
\def\del {\partial}

\def\bZ{\mathbb{Z}}
\def\bR {\mathbb{R}}

\def\calO {{\cal O}}

\def\calM {{\cal M}}

\def\calB {{\cal B}}

\def\bZ {\mathbb{Z}}

\def\ge{\geqslant}
\def\le{\leqslant}

\def\leq{\leqslant}
\def\<{\langle}
\def\>{\rangle}
\newcommand{\diffop}[2]{\ifthenelse{\equal{#2}{1}}{\frac{\mrm{d}}{\mrm{d} #1}}{\frac{\mrm{d}^#2}{\mrm{d} #1^#2}}}

\newcommand{\be}{\begin{equation}}
\newcommand{\ee}{\end{equation}}
\newcommand{\bea}{\begin{eqnarray}}
\newcommand{\eea}{\end{eqnarray}}

\newcommand{\mrm}[1]{{\mathrm #1}}

\def\kappa{\varkappa}
\def\tr{\mathrm{tr}}

\allowdisplaybreaks

\date{October 2018}

\preprint{CERN-TH-2018-225}

\title{\LARGE General Properties of Multiscalar RG Flows in $d=4-\veps$}

\author{Slava Rychkov$^{a,b}$ and Andreas Stergiou$^{c,d}$}

\affiliation{$^a$Institut des Hautes \'Etudes Scientifiques,
Bures-sur-Yvette, France\\
$^b$Laboratoire de physique th\'eorique, D\'epartement de physique de
l'ENS, \'Ecole normale sup\'erieure,\\
\vspace{-2pt}PSL University, Sorbonne Universit\'es, UPMC Univ.\ Paris 06, CNRS, 75005 Paris, France\\
$^c$Theoretical Physics Department, CERN, 1211 Geneva 23, Switzerland\\
$^d$Theoretical Division, MS B285, Los Alamos National Laboratory, Los
Alamos, NM 87545, USA}

\abstract{Fixed points of scalar field theories with quartic interactions
in $d=4-\veps$ dimensions are considered in full generality. For such
theories it is known that there exists a scalar function $A$ of the
couplings through which the leading-order beta-function can be expressed as
a gradient. It is here proved that the fixed-point value of $A$ is bounded
from below by a simple expression linear in the dimension of the vector
order parameter, $N$. Saturation of the bound requires a marginal deformation, and is shown to arise when fixed
points with the same global symmetry coincide in coupling space.  Several
general results about scalar CFTs are discussed, and a review of known
fixed points is given.  \\
\begin{center} \emph{Dedicated to the memory of Louis Michel (1923-1999),\\\vspace{-2pt} the
first IHES professor of physics and
a pioneer of group theory applications to
RG flows.}
\end{center}
}

\begin{document}

\maketitle

\toc

\newsec{Introduction}
Renormalization group (RG) flows in scalar field theories have connections
to innumerable problems in physics.  One is usually interested in
properties of these flows and their fixed points in physical dimension
$d=3$, and the classic approach to learn about such fixed points is to
analytically continue from $d=4-\veps$ dimensions \cite{Wilson:1971dc}.  In
this paper we would like to consider the general case of the Wilson--Fisher
RG equation for $N$ real scalar fields in $d=4-\veps$ dimensions with the
general quartic self-interaction
\eqn{\tfrac{1}{4!}\lambda_{ijkl}\lsp
\phi_i\phi_j\phi_k\phi_l\,}[]
with real symmetric tensor $\lambda_{ijkl}$. The one-loop beta-function has
the well-known form\footnote{Here $t=\ln(\mu/\mu_0)$, with $\mu$ the RG scale, is the RG time. In this paper we consider RG flows from UV to IR, i.e.~$t$ is decreasing along the flow.}
\eqn{\beta_{ijkl}=\frac{d}{dt}\lambda_{ijkl}=-\veps\lsp\lambda_{ijkl}+
\Bconst(\lambda_{ijmn}\lambda_{mnkl}+\text{2 permutations})\,,}[betaOneLoopIntro]
 where $\Bconst=1/{16\pi^2}$ in the standard normalization. From now on we will rescale the coupling so that $B=1$ in the beta-function.

 We will be studying the beta-function equation in the
shown one-loop approximation.  Ideally we would like to get a global
picture of fixed points and RG flows described by this equation. Not much
is actually known about this problem in full generality. As we will review
below, a full classification of fixed points without any assumptions is
available only for $N=1,2$. The problem of classifying fixed points can be
seen as a difficult problem of real algebraic geometry.

In this paper our goal will be to review what is known about multiscalar
fixed points, and to offer a few new general results which may guide future
work towards full classification. Except for a few comments in section
\ref{sec:zero}, we limit ourselves to the one-loop case as it is already sufficiently nontrivial.

In section \ref{Stab} we analyze how stability of the quartic potential  changes under RG flows. We show that fixed points have stable potential. We also show that while a stable potential may become unstable under RG flow, the inverse never happens.

In section \ref{Review} we give a representative review of many known classes of fixed points. This section also reviews a classic construction of fixed points with symmetries possessing one quadratic and two quartic invariants.

In section \ref{Afunction} we recall that the multiscalar RG flow is a gradient flow and present the corresponding height function, the $A$-function.
Since the $A$-function decreases monotonically under RG flows, it's clearly of interest to know what is the minimal value it can take at a fixed point. We prove such a general bound, which scales linearly in $N$, in section \ref{Abound}.
We show by examples that for almost all $N$, and in particular for all $N\ge 12$, our bound is best possible.

In section \ref{sec:RGstab} we study RG stability of fixed points. We review general results about uniqueness of RG stable fixed points, and symmetry criteria for RG instability, following mostly the work of L.~Michel. We also discuss and resolve a paradox of spurious zero eigenvalues of linearized RG equations.

Our main new result is the general bound on the $A$-function. We hope that this bound as well as our review of other existing general results will stimulate further work on general theory of multiscalar fixed points.

\newsec{Stability of the potential}
\label{Stab}
Physically, one is mostly interested in quartic couplings $\lambda_{ijkl}$
such that the potential is stable, which means
that~\cite{Brezin:1973jt}
\eqn{\lambda(\phi) \equiv \lambda_{ijkl}\lsp
  \phi_i\phi_j\phi_k\phi_l\ge 0\qquad
  \text{for any real $\phi_i$}\,.}[stab]
We will call $\lambda$ satisfying this condition ``potential-stable" or simply ``stable'' (this should
not be confused with RG-stability of RG fixed points, to be discussed in section \ref{sec:RGstab}).\footnote{Condition \stab for four-tensors may
be seen as a generalization of the condition for a symmetric matrix to be positive semidefinite. However it's quite more subtle than for matrices. For example checking this condition for a general four-tensor is NP-hard. Also, it's not true that a stable four-tensor can be written as a positive linear combination of elementary tensors $y_i y_j y_k y_l$ for different $y\in \bR^N$ (even allowing for infinite combinations). See \cite{Eng1,Eng2}.}

The set \cC of stable tensors $\lambda_{ijkl}$ forms a convex
cone, which means that (a) if $\lambda\in\cC$, then its rescaling
$s\lsp\lambda \in\cC$ for any $s\ge 0$, and (b) if two tensors
$\lambda^{(1)}$ and $\lambda^{(2)}$ are in $\cC$, then so are their convex linear
combinations:
\eqn{s\lsp\lambda^{(1)}+ (1-s)\lambda^{(2)} \in\cC\qquad
\text{for any } 0\le s\le 1\,.}[]
We will refer to $\cC$ as the stability cone.

We now wish to study RG trajectories which start in the complement of \cC,
so we pick a point $\lambda_{0,ijkl}$ in coupling space which is not in
\cC. This means that there exists a real $\bar{\phi}_i$ such that
$\lambda_0(\bar{\phi})<0$. Let us do an infinitesimal RG flow step to the
IR,
\eqn{\lambda_0 \to \lambda = \lambda_0 +\Delta t\,\beta(\lambda_0)\,,\quad
\Delta t<0\,,}[]
and evaluate the quartic potential \emph{on the same field configuration},
i.e.\ $\lambda(\bar{\phi})$. Using the form of the beta-function, we find that $\lambda(\phi)$ evolves according to
\eqn{
	\frac{d}{d(-t)}\lambda(\bar{\phi}) = \veps\lsp
    \lambda(\bar{\phi})-3\lsp V_{ij} V_{ij} \le
	\veps\lsp \lambda(\bar{\phi}),}[onestep]
where $V_{ij}=\lambda_{ijmn}\bar\phi_m\bar\phi_n$ and we used the fact $V_{ij}V_{ij}\ge 0$.
From the form of this equation we see that if $\lambda_0(\bar \phi)<0$ then
the right-hand side is always negative and
so the potential evaluated on the field configuration $\bar\phi$ gets more
and more negative as the flow towards the IR progresses \cite{Brezin:1973jt}. This has two consequences.
First, the RG flow remains in the complement of \cC if it starts there: RG
flows cannot enter the stability cone. Second, there cannot be any fixed
point in the complement of \cC.\footnote{This second observation can also be
shown directly from the beta-function equation, which implies
$\lambda_*(\phi)=3\veps\lsp B\lsp V_{ij}V_{ij}\ge0$ for a fixed point $\lambda_*$ \cite{Michel:1983in}.}

Notice, however, that RG flows can \emph{exit} the stability cone, as
explicit examples show. A well known example occurs in theory of
the cubic anisotropy as discussed e.g.\ in~\cite[sec.\
11.3]{Pelissetto:2000ek}. Such RG flows are known as ``fluctuation driven first-order phase transitions".

\begin{figure}[t!]
	\centering
	\includegraphics[width=0.8\linewidth]{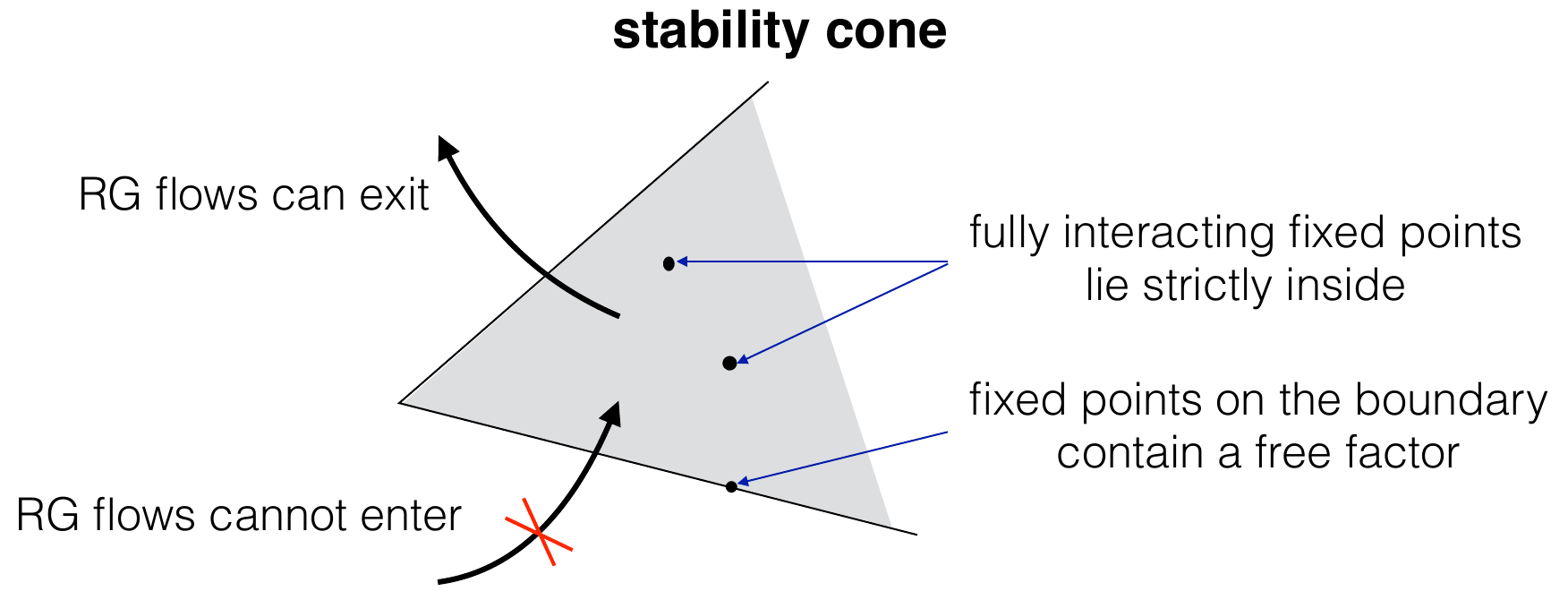}
	\caption{\label{fig-stability} Graphical summary of the results of section \ref{Stab}.}
\end{figure}

Can there be fixed points on the boundary of the stability cone?  Being on
the boundary means that there is a \emph{flat} direction $\bar \phi_i$ in
field space, $\lambda(\bar \phi)=0$, while in all other directions
the potential is non-negative.

Let us introduce some terminology. The fixed point
$\lambda=0$ is called \emph{trivial} or \emph{free}.  Two fixed point tensors $\lambda$ and
$\tilde\lambda$ that can be transformed into one another by an $O(N)$
transformation (change of basis) of course describe physically equivalent fixed points. If we
can split the fields, perhaps after a change of basis, into two
subsets so that only fields within each group interact with each other,
such a fixed point is called \emph{factorized}.  Finally, a fixed point
which cannot be factorized is called \emph{fully interacting}.

We will now show that all fixed points on the boundary of the stability cone are
either free, or contain a free factor.  This will imply that all
fully interacting fixed points lie strictly inside the stability cone: \stab
is strictly positive for all nonzero $\phi$; see Fig.~\ref{fig-stability}.

Let $\lambda$ be a fixed point on the boundary of the stability cone, and
$\bar \phi$ be a flat direction.  Rotating fields we can assume that
$\bar \phi=(1,0,0,\ldots)$ points in direction 1. Then,
$\lambda(\bar\phi)=0$ means that $\lambda_{1111}=0$. We would like to
show that all other couplings involving at least one index 1 vanish, so
that the $\phi_1$ subsector is completely free. To do this we use
the fixed point condition. Using that the beta-function for
$\lambda_{1111}$ vanishes, we obtain
\eqn{0=\beta_{1111}=-\veps\lsp\lambda_{1111} + 3\lambda_{11mn}
\lambda_{11mn} = 3\lambda_{11mn} \lambda_{11mn}\,,}[]
from where we conclude that all couplings $\lambda_{11mn}$ vanish. Now we use the beta-function equation for $\lambda_{11jj}$ where $j$ is an
arbitrary index (no sum on $j$):
\eqn{0=\beta_{11jj}=-\veps\lsp\lambda_{11jj}+
(\lambda_{11mn}\lambda_{jjmn}+2\lsp\lambda_{1jmn}\lambda_{1jmn})
= 2 \lambda_{1jmn}\lambda_{1jmn}\,,}[]
where we used that all $\lambda_{11mn}$ were already proved to vanish.
Therefore, $\lambda_{1jmn}=0$ for any $j,m,n$, which is what we need.

\newsec{Review of known fixed points}\label{Review}

In this section we will attempt a review of known fixed points. Naturally
we will only mention fully interacting fixed points, as defined in section
\ref{Stab}. A fundamental characteristic of any fixed point is its symmetry
group $G$, which is defined as the \emph{maximal} subgroup of $O(N)$ that
leaves the tensor $\lambda_{ijkl}$ invariant.

Notice that if $G\ne O(N)$, then by applying an $O(N)$ transformation not
in $G$ the tensor $\lambda_{ijkl}$ is transformed to a different tensor
$\tilde\lambda_{ijkl}$, which nevertheless describes the same physics.
Classifying fixed points means classifying solutions of the beta-function
equation $\beta_{ijkl}=0$ up to this equivalence relation. However, there
is usually one choice of field basis where the fixed point tensor
$\lambda$ takes a particular simple form.

The symmetry group of any fixed point is at least as large as $\bZ_2$,
which acts by simultaneous sign flips on all  fields. Curiously, all known
fixed points for $N\ge 2$ have a strictly larger symmetry group.  It would
be interesting to understand why it is so.\\[5pt]

{\bf Open problem.}\footnote{A bottle of Dom
	P\'erignon champagne will be awarded for a solution of this problem. Please contact the authors for collecting the prize.} Construct a fully
	interacting $N\ge 2$ scalar one-loop fixed
	point in $4-\veps$ dimensions with real couplings and just $\bZ_2$ symmetry, or prove that all such fixed points have strictly larger symmetry. \\[5pt]

An important characteristic of the symmetry group $G$ are the numbers $I_2$
and $I_4$ of quadratic and quartic invariants, $A_{ij}\phi_i \phi_j$ and $
B_{ijkl} \phi_i\phi_j\phi_k\phi_l$, where $A_{ij}$ and $B_{ijkl}$ are
linearly independent symmetric two- and four-tensors invariant under $G$
(to count them we choose and fix a basis in field space). When $G=O(N)$
there is just one quadratic, $\vec\phi^{\lsp 2}$, and one quartic,
$(\vec\phi^{\lsp 2})^2$, invariant, and the question is if there are more
when the symmetry is reduced.  The number of quartic invariants is clearly
important since the fixed point tensor $\lambda$ will be a linear
combination of $I_4$ independent invariant tensors.

While quadratic invariants do not enter directly into the analysis of RG
equations, their number is important for the physical interpretation of the
fixed points.  Terms quadratic in fields are strongly relevant
perturbations of the potential. $G$-noninvariant quadratic terms are
forbidden by symmetry, while all $G$-invariant quadratic terms have to be
fine-tuned to zero to reach the fixed point. Groups $G$ for which
$\vec\phi^{\lsp 2}$ remains a single quadratic invariant are particularly
interesting, since fixed points with such symmetry would require less
fine-tuning to be realized in an experiment.\footnote{Notice that some of
	the quartic perturbations may also be relevant, but those require
	additional analysis. See section \ref{sec:RGstab}.} A single quadratic invariant ($I_2=1$) is equivalent
to requiring that the fundamental representation of $O(N)$ remains
irreducible under $G$.

Historically, most attention was dedicated to fixed points with $I_2=1$.
Notice, however, that a full classification requires considering fixed
points that do not necessarily satisfy this condition.\footnote{Ref.~\cite[sec.~3]{Michel} contains a remark which seems to suggest that fixed points with $I_2>1$ can always be factorized into a product of pairwise noninteracting fixed points with $I_2=1$. This cannot be correct as the example of biconical fixed point below shows. Fortunately this remark is quite tangential in \cite{Michel} and does not affect the validity of the main considerations.}
We will now give some prominent examples of families of fixed points (see Table
\ref{tab:FPex}), known to exist for infinitely many values of $N$. Some of
them have a discrete and some a continuous symmetry group. Our list is
representative but far from complete; see e.g.~\cite{Vicari:2007ma,Osborn:2017ucf} for more
examples.  We will then discuss what is known about classification.

\begin{table}
	\centering
	\begin{tabular}{lllll}
		\toprule
		Name 	& $N$ 		& $G$ 								& $I_4$ & $I_2$ \\
		\midrule
		$O(N)$ 	& $N\ge 1$ & $O(N)$ 						& 1 		& 1 	\\
		cubic  	& $N\ge 3$ & $ (\bZ_2)^N \rtimes S_N$ & 2 & 1 \\
		tetrahedral & $N\ge 4$ & $S_{N+1}\times\bZ_2$  & 2 & 1\\
		bifundamental & $N=mn$ & $O(m)\times O(n)/\bZ_2$ & 2 & 1 \\
		& $ (m,n\ge 2, R_{mn}\ge 0 )$ &\\
		``MN"		& $N=mn$ & $O(m)^n\rtimes S_n$ & 2 & 1\\
		& $(m,n\ge 2, m\ne 4)$ & \\
		tetragonal & $N=2n\ge 4$ & $ (D_8)^n \rtimes S_n$ & 3 & 1\\
		Michel & $N=r_1\cdots r_k$ & $ G_{r_1\ldots r_k} $ & $k+1$ & 1\\
		biconical\tablefootnote{For $m_1=m_2=m$, this fixed point is a particular case of the MN fixed point with $n=2$, and the symmetry is enhanced to $O(m)^2\rtimes \bZ_2$, reducing the number of quadratic invariants to 1.} & $N=m_1+m_2$ & $O(m_1)\times O(m_2)$
		& 3& 2\\
		\bottomrule
	\end{tabular}
	\caption{Summary of examples of fully interacting fixed points given in text.}
	\label{tab:FPex}
\end{table}


Maximal symmetry $G=O(N)$ is realized for the {\bf O(N) fixed point} with
quartic potential given by $\lambda (\vec\phi^{\lsp 2})^2\,$. It exists
for any $N\ge 1$, and reduces for $N=1$ to the Ising (also called Wilson--Fisher) fixed point.

\subsec{Fixed points with \texorpdfstring{$I_2=1$}{I\_2=1},
\texorpdfstring{$I_4=2$}{I\_4=2}: general theory}
\label{sec:I42gen}
We next consider fixed point symmetries which allow two quartic and one quadratic
invariant. There is a neat general theory of such fixed points, which we will now review.
First of all they satisfy the famous trace condition of \cite{Brezin:1973jt}:
\eqn{\lambda_{iijk}=\veps\lsp z\lsp \delta_{jk}\,,}[satTrace1]
where $z$ is some (fixed-point dependent) constant. Indeed, the trace in the left-hand side is a $G$-invariant two-tensor and
since $I_2=1$ it must be proportional to the tensor $\delta_{jk}$. It is
then natural to write $\lambda$ as a sum of two terms:
\eqn{\lambda_{ijkl}=\veps\lsp\big(\tfrac{1}{N+2} z \lsp T_{ijkl}+
  d_{ijkl}\big)\,,\qquad
	T_{ijkl}=\delta_{ij}\delta_{kl}+\delta_{ik}\delta_{jl}
	+\delta_{il}\delta_{jk}\,.}[lamAns01]
As a consequence of \eqref{satTrace1}, the tensor $d_{ijkl}$ defined by this equation will be symmetric and traceless.

We now impose that the coupling \eqref{lamAns01} satisfies the
beta-function equation. Using the fact that $d_{ijkl}$ is symmetric and traceless, the beta-function equation reduces to
\eqn{d_{ijmn}d_{klmn}+d_{ikmn}d_{jlmn}+d_{ilmn}d_{jkmn}= p\, T_{ijkl}+
	q\lsp d_{ijkl}\,,}[ddpermsContr1]
with coefficients $p, q$ given by
\eqn{p=\frac{1}{N+2}\lsp z\Big(1-\frac{N+8}{N+2}\lsp z\Big)\,,\qquad
q= 1 - \frac{12}{N+2}\lsp z\,.}[pqBound1]
Notice that Eqs.~\eqref{lamAns01}, \eqref{ddpermsContr1} and \eqref{pqBound1} followed from the trace condition only. This will be useful in section \ref{sec:sat}.

Now we will use the assumption $I_4=2$, i.e.~that the space of invariant symmetric
four-tensors is two-dimensional. We take as its basis elements $T_{ijkl}$ and another tensor $d^{\llsp G}$, chosen traceless without loss
of generality.\footnote{This
	additional tensor $d^{\llsp G}$ is called ``primitive", because by
	assumption it cannot be reduced to products of lower-rank invariant
  tensors.} Then, the tensor $d$ in \eqref{lamAns01} must be proportional
    to $d^{\llsp G}$: $d=\alpha d^{\llsp G}$.
Notice that the tensor $d=\lsp d^{\llsp G}$ is bound to satisfy
Eq.~\eqref{ddpermsContr1} with some $p=p_G$, $q=q_G$. Indeed, the left-hand
side of \eqref{ddpermsContr1} is a $G$-invariant symmetric four-tensor, so it must be expressible as a linear combination of $T$ and $d^{\llsp G}$. Notice also that \eqref{ddpermsContr1} implies
\eqn{d_{ijkl}d_{ijkl}=\tfrac12\llsp p N(N+2) ,}[ddSq1]
and so $p_G> 0$ since we assume that $d^{\llsp G}$ is not identically vanishing.

The tensor $d=\alpha\lsp d^{\llsp G}$ will then satisfy
Eq.~\eqref{ddpermsContr1} with $p=\alpha^2\llsp p_G$, $q=\alpha\lsp q_G$.
Substituting these into \eqref{pqBound1} we see that in order to find the
fixed point we must solve
\eqn{\alpha^2\llsp p_G=\frac{1}{N+2}\lsp z\Big(1-
	\frac{N+8}{N+2}\lsp z\Big)\,,\qquad
  \alpha\lsp q_G=1 -\frac{12}{N+2}\lsp z\,,}[FPz1]
for $\alpha$ and $z$.  Since $p_G\ge 0$ we find $0\le z\le
\frac{N+2}{N+8}$.

There are two solutions of \eqref{FPz1}, $(\alpha_+, z_+)$ and $(\alpha_-,z_-)$,
with
\eqn{\alpha_{\pm}=\frac{1}{q_G}\frac{(N+2)\rho\pm 6 \lsp\sqrt{\Delta}}
	{144+(N+8)\rho}\,,\qquad
    z_\pm=\frac{(N+2)(24+\lsp \rho \mp \sqrt{\Delta})}
	{2\big(144+(N+8)\rho\big)}\,,}[]
where $\rho =q_G^{\lsp 2}/p_G\ge0$ and
\eqn{\Delta=\rho\big(\rho-4(N-4)\big)\,.}[]
If $\rho\ge 4(N-4)$ we have a pair of fixed points with real couplings, which coincide for $\rho=4(N-4)$.
If $\rho<4(N-4)$, these fixed points have complex couplings and are
discarded since here we are interested in real fixed
points.\footnote{Although in other contexts fixed points with complex
couplings may be useful, see \cite{Gorbenko:2018ncu,Gorbenko:2018dtm}.} A
related discussion about the presence of pairs of fixed points has
appeared in~\cite{Brezin:1973jt, Osborn:2017ucf}.\footnote{More generally, Ref.~\cite{Brezin:1973jt} shows that given any fixed point $\lambda_*$ satisfying the trace condition, whatever its symmetry, there is another fixed point $\lambda_{**}$ which can be expressed as a linear combination of $T$ and $\lambda_*$.}

Fixed points obtained using this construction are usually fully
interacting, but sometimes they factorize. This happens for one of the
cubic symmetry fixed points, and for one of the MN fixed points; see below.

\subsec{Fixed points with \texorpdfstring{$I_2=1$}{I\_2=1},
\texorpdfstring{$I_4=2$}{I\_4=2}: examples}
\label{sec:I42ex}
We will now consider several examples where the above general theory can be applied.

The {\bf cubic fixed point} has $G= (\bZ_2)^N \rtimes S_N$
symmetry, called the cubic group \cite{Aharony, Aharony2, Wallace2,
Osborn:2017ucf}. This is the symmetry group of the unit cube in $N$
dimensions. The second quartic invariant is $\sum_i \phi_i^4$, in
the frame in which $G$ acts by permuting the fields, and by flipping their
signs. Forming the $d^{\llsp G}$ tensor for cubic symmetry and computing the $\rho$ parameter we find
\eqn{\rho_C=\frac{9(N-2)^2}{2(N-1)}\qquad (N\ge 3)\,.}[rhoC1]
We see that $\rho_C>4(N-4)$ for all $N\ge 3$, and so for any $N\ge 3$
there are two fixed points with this symmetry. One
is fully interacting, while the other consists of $N$ decoupled copies of
the Ising fixed point.

The {\bf tetrahedral fixed points} have $G=S_{N+1}\times\bZ_2$, the tetrahedral
group~\cite{ZiaW, Osborn:2017ucf}. Consider vectors $e_i^\alpha$,
$\alpha=1,\ldots,N+1$ satisfying
\beq
\sum_\alpha e_i^\alpha = 0,\quad e^\alpha_i e^\beta_i = \delta^{\alpha\beta}-\frac{1}{N+1}\,,
\eeq
which are vertices of the perfect hypertetrahedron in $\bR^N$.
The $S_{N+1}$ part of $G$ is the symmetry group of this hypertetrahedron,
permuting the vertices, and $\bZ_2$ acts by flipping the sign of all
fields. The second quartic invariant involves the tensor $\sum_\alpha e^\alpha_i e^\alpha_j e^\alpha_k e^\alpha_l$. This is sometimes referred to as the
restricted Potts model.
The $\rho$ parameter for the tetrahedral symmetry equals
\eqn{\rho_T=\frac{9(N^2-3N-2)^2}{2(N-2)(N-1)(N+1)}\qquad (N\ge3)\,.}[]
We can check that $\rho_T\ge4(N-4)$ for all $N\ge 3$,
with a strict inequality for all $N$ except for $N=5$ where it's an
equality. There are therefore two fixed points with this symmetry for $N\ge 4$, which coincide for $N=5$. For $N=3$ the tetrahedral group $G$ is isomorphic to the cubic group, and the tetrahedral fixed points coincide with the cubic ones (indeed $\rho_C=\rho_T=\frac94$ for $N=3$).

The {\bf bifundamental fixed points}\footnote{This is our proposed terminology. Sometimes they are called $O(m)\times O(n)$ fixed points.} have $G=O(m)\times O(n)/\bZ_2$, where
$N=mn$~\cite{Kawamura, Pelissetto:2001fi, Vicari:2007ma,Osborn:2017ucf}. We restrict to $m\ge n$ without loss of generality. To realize
these fixed points, one writes $\phi_i$ as a $m\times n$ matrix field $\Phi_{ab}$,
which transforms as a bifundamental of $O(m)\times O(n)$. The second quartic invariant is $\tr (\Phi \Phi^T \Phi \Phi^T)$.
Since both $O(m)$ and $O(n)$ can realize the same overall sign flip
$\Phi\to -\Phi$ we need to mod out by a $\bZ_2$.
\begin{figure}[t!]
	\centering
	\includegraphics[width=0.35\linewidth]{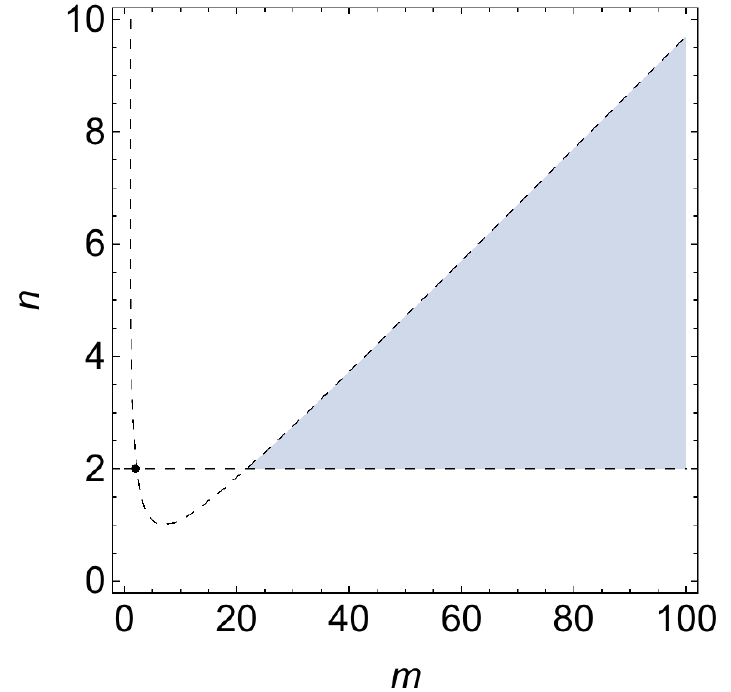}
	\caption{\label{fig-bifund} The region of $(m,n)$ space satisfying the conditions $m\ge n\ge 2$, $R_{mn}\ge0$ needed to have bifundamental fixed points with real couplings. It consists of the point $m=n=2$, and of the integer points in the gray region, described by Eq.~\eqref{eqnlessm}. Notice that the allowed region around point $m=n=2$ is really a tiny triangle invisible on this scale, but $m=n=2$ is the only integer point within it.}
\end{figure}

Using results
of~\cite{Osborn:2017ucf} we find the $\rho$ parameter
\begin{align}
\rho_{\text{bif}}&=\frac{\big(mn(m+n)+4\lsp
	mn-10(m+n)-4\big)^2}{3(m-1)(m+2)(n-1)(n+2)}\\
&= 4(N-4)+\frac{(mn+2)^2\lsp R_{mn}}
{3(m-1)(m+2)(n-1)(n+2)}\,,
\end{align}
where
\eqn{
	R_{mn}=m^2+n^2-10\lsp mn-4(m+n)+52\,.}[Rmn]
To have fixed points with real couplings we need to impose $R_{mn}\ge 0$. Restricting to $m\ge n$, this is satisfied by $m=n=2$ and
\eqn{ 2\le n\le 5m+2-2\sqrt{6(m+2)(m-1)}\qquad (m\ge 22),}[eqnlessm]
see Fig.~\ref{fig-bifund}.
If $R_{mn}>0$ there are two fixed points with real couplings (which are sometimes referred to as the chiral and antichiral fixed points), which coincide if $R_{mn}=0$.

The Diophantine equation $R_{mn}=0$ has an infinite number of positive
integer solutions
given by~\cite{Dario}
\eqna{m_i&=10\lsp m_{i-1}- n_{i-1}+4\,,\quad n_i=m_{i-1}\,,
  \quad i=1,2,\ldots,\\
	m_1&=7\,,\quad n_1=1\,.}[Dioph]
Since $n>1$ the solution with smallest $N$ is $m=73$, $n=7$, $N=511$.

We finally consider the {\bf MN fixed points}. They have $G=O(m)^n\rtimes S_n$, where again $N=mn$ but in this case there is no symmetry between $m$ and $n$~\cite{Mukamel2, Shpot,
Shpot2, Mudrov, Michel, Osborn:2017ucf}. In this case $\phi_i$ is
decomposed as $n\ge 2$ vectors of size $m$, $\vec{\varphi}_r, r=1,\ldots,n$,
and the second quartic invariant is $\sum_{r}(\vec{\varphi}_r^{\lsp
	2})^2$. The case $m=1$ is equivalent to the cubic, and the case $m=n=2$ to the bifundamental
symmetry.\footnote{As a check, $\rho_{\text{MN}}$ coincides with the cubic $\rho_C$ for $m=1$, $n\ge 2$, and with the bifundamental $\rho_{\text{bif}}$ for $m=n=2$.}

Using again results of~\cite{Osborn:2017ucf}, the $\rho$ parameter is given by
\begin{align}\rho_{\text{MN}}&=\frac{\big(m(m+8)(n-1)+(m-4)(m+2)\big)^2}
{6m(m+2)(n-1)}\\
&= 4(N-4)+\frac{(m-4)^2(mn+2)^2}{6\lsp m(m+2)(n-1)}\,.
\end{align}
General theory predicts that there are two real fixed points for $m\ne 4$, which coincide for $m=4$.
However, only one of these fixed points is fully interacting, while the other is factorized, consisting of $n$ decoupled $O(m)$ theories. For $m=4$ only the factorized fixed point remains.

\subsec{Further examples of fixed points}
\label{sec:further}

As an example of a fixed point with three quartic invariants and one quadratic invariant, we mention the {\bf tetragonal fixed
	point}~\cite{Mukamel2, Mukamel3, Mudrov2, Osborn:2017ucf},\cite[sec.~11.6]{Pelissetto:2000ek}, which exists for even $N=2n\ge 4$. The quartic potential includes the isotropic $O(N)$ term,
the cubic term $\sum_i \phi_i^4$, and the tetragonal anisotropy. The latter
takes the form
$\phi_{1}^2\phi_{2}^2+\phi_3^2\phi_4^2+\ldots +\phi_{N-1}^2\phi_N^2$.
Focussing on the first pair of fields $\phi_1,\phi_2$, permuting them and flipping signs generates the eight-element dihedral group $D_8$. The full symmetry group of this fixed point is therefore $(D_8)^{n}\rtimes S_{n}$.

There exist fixed points with arbitrary large $I_4$. An example is provided
by the {\bf Michel fixed point}~\cite{Michel, GrinsteinL, Osborn:2017ucf}, which is a generalization of the MN fixed point. Consider $N=r_1\cdots r_k$ where $r_i>1$ are integers (not necessarily prime), and write $\phi_i$ as a tensor with $k$ indices $\alpha_i=1\ldots r_i$, $\Phi_{\alpha_1\ldots \alpha_k}$. The quartic term
\beq
\sum_{\alpha_1\ldots \alpha_l}\left(\sum_{\alpha_{l+1}\ldots \alpha_k}\Phi^2_{\alpha_{1}\ldots \alpha_k}\right)^2
\eeq
breaks $O(N)$ to $O(m)^n\rtimes S_n$ where $n=r_1\cdots r_l$ and $m=r_{l+1}\cdots r_k$.
The Michel fixed point contains $k+1$ such terms ($0\le l\le k$) with nonzero couplings, so that the symmetry group $G_{r_1\ldots r_k}$ is the intersection. 

Finally we mention the {\bf biconical fixed point}, which provides an example with more than one
quadratic invariant. Let us split $\vec \phi$ into two vectors $\vec
\phi_1$ and $\vec\phi_2$ with $m_1$ and $m_2$ components, $m_1+m_2=N$, and
consider the symmetry group $O(m_1)\times O(m_2)$ acting on $\vec \phi_1$
and $\vec\phi_2$. There are two quadratic invariants, $\vec \phi_1^{\lsp 2}$
and $\vec\phi_2^{\lsp 2}$, and the quartic potential is a linear combination
of three invariants, $(\vec \phi_1^{\lsp 2})^2$, $(\vec\phi_2^{\lsp 2})^2$,
and $\vec \phi_1^{\lsp 2} \vec\phi_2^{\lsp 2}$
\cite{Nelson:1974xnq,Kosterlitz:1976zza,Calabrese:2002bm}.

\subsec{Classification results}
\label{sec:class}

Full classification of fixed points is available only for $N=1$ and $N=2$. Namely, for $N=1$ we have only two fixed points, both with $G=\bZ_2$: the free one at $\lambda=0$ and the Wilson--Fisher fixed point at $\lambda=\veps/3$.

For $N=2$ we have only one fully interacting fixed point: the $O(2)$ one;
see \cite{Osborn:2017ucf} for a completely general proof.

For $N=3$ there are three known fully interacting fixed points: $O(3)$, cubic, and biconical. The $O(3)$ and cubic fixed points are the only ones assuming the ``single quadratic invariant"
condition (or a more general ``isotropy constraint''
$\lambda_{iklm}\lambda_{jklm}\propto \delta_{ij}$ \cite{Zia:1974nv}). The biconical fixed point has $O(2)\times \bZ_2$ symmetry \cite{Nelson:1974xnq,Kosterlitz:1976zza,Calabrese:2002bm}.\footnote{We thank Matthijs Hogervorst for reminding us about the $N=3$ biconical fixed point.} It would be nice to prove rigorously that there are no further fixed points.

Under the assumption of a single quadratic invariant, an extensive analysis of fixed points was performed
in~\cite{Brezin2} for $N=4$, and in~\cite{Hatch, Hatch1,
	Hatch2} for $N=6$. These works identified dozens of fixed points,
    corresponding to various discrete subgroups of $O(4)$ and $O(6)$,
    respectively. They offer a glimpse of
    the incredible complexity that a full classification of fixed points is bound to entail.

\section{The \texorpdfstring{$\boldsymbol{A}$}{A}-function}
\label{Afunction}
As we have seen in the previous section, there are many fixed points. Ideally we would like to understand all fixed points and RG flows connecting them. There are currently only partial results towards this goal.

It has been observed long ago by Wallace and Zia \cite{Wallace:1974dx, Wallace:1974dy} that
the one-loop beta-function can be written as a gradient of an
$A$-function:
\eqn{\beta_{ijkl}=\frac{\delta}{\delta\lambda_{ijkl}}A\,,\qquad
A=-\tfrac12\lsp\veps\lsp\lambda_{ijkl}\lambda_{ijkl}
+\lambda_{ijkl}\lambda_{klmn}\lambda_{mnij}\,.}[Acubic]
We use the variational and not the usual derivative in \Acubic because couplings
are real symmetric rank-four tensors, so the components $\lambda_{ijkl}$
are not all independent. Eq.~\Acubic thus means that the variation of $A$
is expressible as
\eqn{
	\delta A = \beta_{ijkl}\lsp\delta\lambda_{ijkl}\,.
}[AcubicGR]
This is the same convention as when varying with respect to the metric in general relativity.

Equivalently we can consider a bigger vector space of all real rank-four
tensors, call it $V_4$, of which the vector space of symmetric couplings,
$V_4^{\rm sym}$, is a subspace. The $A$-function can be formally considered
as given by the same equation on the full $V_4$. The variational derivative
in \Acubic can be computed as the usual partial derivative
$\frac{\partial}{\partial\lambda_{ijkl}}$ applied to the so-extended
$A$-function.

It will also be helpful to write Eq.~\Acubic in a form which refers to an
independent set of coordinates on $V_4^{\rm sym}$. Let $\lambda^I$ be such
a set of coordinates and $g_{IJ}$ be the restriction of the flat metric on
$V_4$ to $V_4^{\rm sym}$.\footnote{For example we can choose as $I$ ordered
tuples $ijkl$.  It's easy to see that with this choice
$g_{IJ}=p_I\delta_{IJ}$ where $p_I$ is the number of non-identical
permutations of the tuple $I$. E.g.~$p_{1111}=1$, $p_{1112}=4$, etc.} Then
\Acubic can be equivalently expressed as\footnote{We will write
$\lambda^I$ and $\beta^I$ as required by the differential geometry
conventions on contravariant and covariant indices. However, we will keep
lower indices in $\lambda_{ijkl}$ and $\beta_{ijkl}$. Hopefully this will
not cause confusion.}
\eqn{\beta^I = g^{IJ}\del_J A\,.}[Acubiccov]

Eq.~\AcubicGR or its more covariant form \Acubiccov imply that the
$A$-function decreases along the RG flow (flowing towards the IR). Indeed
we get $(d/dt)A = \beta^I \del_I A= g^{IJ} \del_I A\, \del_J A \ge 0$.

The existence of the $A$-function plays a fundamental role in the
classification of RG fixed points and of RG flows connecting them.

One useful consequence is as follows. Take an arbitrary RG trajectory. One
possibility is that the trajectory runs out to infinity. Consider the more
interesting possibility that the trajectory stays bounded for all times.
From a general theorem about real-analytic gradient flows due to
\L{}ojasiewicz \cite{Lo,Kurdyka1,LoEnc},
we can conclude:\footnote{Analyticity of $A$ is important. For example, one can construct a $C^\infty$ gradient flow with a trajectory whose limit set is not a single point but a segment. For real-analytic $A$ such pathologies are impossible.}

{\bf Fact.} Any bounded RG trajectory necessarily goes to a fixed point.

Of particular interest is the value $A_\ast$ of $A$ at the fixed point.
Contracting the beta-function equation $\beta_{ijkl}=0$ with
$\lambda_{ijkl}$ we have
\eqn{\veps\lsp \lambda_{*ijkl}\lambda_{*ijkl}
= 3\lsp \lambda_{*ijkl}\lambda_{*klmn}\lambda_{*mnij}\,,
}[betaContrLam]
where $\lambda_{*ijkl}$ stands for a fixed point coupling value. Using this
in equation \Acubic for $A$ we have
\eqn{A_*=-\tfrac16 \lsp\veps\lsp\lambda_{*ijkl}\lambda_{*ijkl}\,,}[AStar]
from where we see that $A_*$ is always negative. It is clearly interesting
to know how negative $A_*$ can become.  One of the main results of our
paper will be to establish a general lower bound:
\eqn{A_\ast\ge-\tfrac{1}{48}\lsp N\lsp\veps^3\,.}[AStarBound] Such a bound
was previously observed in \cite{Osborn:2017ucf} for a class of RG flows
preserving a subgroup of $O(N)$. Here we will show that it is completely
general. In particular, it holds independently of any assumption about the symmetry and the number of quadratic and quartic invariants. It applies both to fully interacting and factorized fixed points.

Equivalently, \AStarBound says that all fixed points belong to a known compact region of coupling space:
\eqn{
	\lambda_{*ijkl}\lambda_{*ijkl} \leq \tfrac{1}{8} N \veps^2\,.}[AStarBoundEq]
Any search of new fixed points can therefore be restricted to this region.

\newsec{A bound on \texorpdfstring{$\boldsymbol{A}$}{A}}
\label{Abound}
In this section we will prove the bound \AStarBound on the value of $A$, or
the equivalent bound \AStarBoundEq, at any fixed point. We will omit the
star subscript for fixed point values, something that will hopefully not
cause any confusions. Just in this section, it will be convenient to further rescale the
couplings $\lambda\to\lambda/\veps$.  Before rescaling the fixed point
coupling is $\text{O}(\veps)$, after rescaling it's $\text{O}(1)$. The
rescaled one-loop fixed point equation takes the form
\eqn{\lambda_{ijkl} = \lambda_{ijmn}\lambda_{mnkl}+\text{2 permutations}\,.
}[betaResc]
We will show that any real symmetric four-tensor solving this equation satisfies the bound
\eqn{S=\lambda_{ijkl}\lambda_{ijkl}\le C_N\,,\qquad C_N=\tfrac{1}{8} N\,.}[ineq1]
This is equivalent to \AStarBoundEq after undoing the rescaling
$\lambda\to\lambda/\veps$.

\subsec{Why a bound is expected to exist}
First we present a simple argument which explains why a bound is expected
to exist. We will fix $N$ and we will try to show that all components of $\lambda_{ijkl}$ are bounded by some constant. Like in an argument seen in section \ref{Stab}, the idea is to first consider components $\lambda_{iiii}$, then $\lambda_{iimn}$, and finally the general case.

For components $\lambda_{iiii}$, considering for definiteness $i=1$, the beta-function \betaResc implies
\eqn{\lambda_{1111}=3\lsp \lambda_{11mn}\lambda_{11mn}\ge
3\lsp\lambda_{1111}^2\,.}[1111beta]
From here we can conclude two things. First, since $\lambda_{1111} \ge
3\lsp\lambda_{1111}^2$, we must have\footnote{Using the notation of section \ref{Stab} this can also be written as $0 \le \lambda(\bar\phi)\le \tfrac 13$ for any unit-length $\bar\phi$ \cite{Brezin:1973jt}.}
\eqn{0\le \lambda_{1111}\le \tfrac 13\,.}[1111]
%
Second, from the first equality in \eqref{1111beta} and from \eqref{1111} we have
\eqn{\lambda_{11mn}\lambda_{11mn} = \tfrac 13\lsp \lambda_{1111}
\le \tfrac 19\,.}[11mn]
In particular, for any $m,n$
\eqn{
|\lambda_{11mn}|\le \tfrac 13\,.}[mnSim]

Finally let us bound components $\lambda_{ijkl}$ where no two indices are equal. Take for definitenes $i=1$, $j=2$, and impose the beta-function
equation \betaResc for $\lambda_{1122}$:
\eqn{\lambda_{1122} = 2\lsp\lambda_{12mn}\lambda_{12mn}
+\lambda_{11mn}\lambda_{22mn}\,.}[]
From here we have
\eqn{2\lsp\lambda_{12mn}\lambda_{12mn} = \lambda_{1122} -
\lambda_{11mn}\lambda_{22mn}\,.}[]
The first term in the right-hand side is bounded by \mnSim, while the
second term can be bounded in absolute value using \eqref{11mn} and
the Cauchy--Schwarz inequality:
\eqn{|\lambda_{11mn}\lambda_{22mn}|\le(\lambda_{11mn}\lambda_{11mn})^{1/2}
(\lambda_{22pq}\lambda_{22pq})^{1/2}\le \tfrac 19\,.}[Cauchy]

Summing the obtained bounds for all components, we will get a bound of the
form \eqref{ineq1} with some constant $C_N$. This reasoning is rather crude
and does not give an optimal constant, in particular $C_N$ will grow
quadratically with $N$ because one will have to sum over all pairs $i,j$
with $i\ne j$. In the next section we will present the argument producing
$C_N=\frac18 N$.

\subsec{The bound with \texorpdfstring{$C_N=\tfrac18 N$}{C\_N=N/8}}
We first introduce some notation. In this section the Einstein summation convention will be applied to indices $m,n$, but all summation in indices $i,j$ will be explicitly indicated.

Denote $x_i=\lambda_{iiii}$. From \eqref{1111} we know that  $x_i\in[0,\frac13]$. Denote also
$(\vv_i)_{mn}=\lambda_{iimn}$, viewed as matrices in $m,n$ indices. Then the first equality in
\eqref{1111beta} can be written as
\eqn{\tr\,\vvi{\llnsp}^2=\tfrac13\lsp x_i\,.}[bouI]
The beta-function equation for the components $\lambda_{iijj}$,
\eqn{\lambda_{iijj}=2\lsp\lambda_{ijmn}\lambda_{ijmn}+\lambda_{iimn}\lambda_{jjmn}\,,}[]
can be written as
\eqn{\lambda_{ijmn}\lambda_{ijmn}=\tfrac12\big((\vvi)_{jj}
-\tr(\vvi\vvj)\big)\,.}[bouII]
The quantity we need to bound takes the form (using \bouI and \bouII)
\begin{align}
S&
=
\sum_i\lambda_{iimn}\lambda_{iimn}
+\sum_{i\ne j}\lambda_{ijmn}\lambda_{ijmn}\\
&=\tfrac13\sum_i x_i+\tfrac12\sum_{i\ne j}(\vvi)_{jj}
-\tfrac12\sum_{i\ne j}\tr(\vvi\vvj)\,.
\label{Sexp2}
\end{align}
Using further the identity
\eqn{\sum_{i\ne j}\tr(\vvi\vvj)=
\tr\Big(\sum_i\vvi\Big)^2-\sum_{i}\tr\,\vvi{\llnsp}^2=
\tr\Big(\sum_i\vvi\Big)^2-\tfrac13\sum_ix_i\,,}[]
we rewrite \eqref{Sexp2} as
\eqn{S=\tfrac12\sum_ix_i+\tfrac12\sum_{i\ne j}(\vvi)_{jj}-
\tfrac12\tr\Big(\sum_i\vvi\Big)^2\,.}[SexpIII]
We estimate the last term in the right-hand side of
\eqref{SexpIII} as follows:
\eqn{\tr\Big(\sum_i\vvi\Big)^2=\sum_{j,k}\Big(\sum_i
(\vvi)_{jk}\Big)^2\ge\sum_j\Big(\sum_i
(\vvi)_{jj}\Big)^2=\sum_i\Big(\sum_j
(\vvi)_{jj}\Big)^2\,,}[keyIneq]
where in the last equality we rename $i\leftrightarrow j$
and use $(\vvi)_{jj}=(\mathbf{v}_j)_{ii}$. We can also separate the $i=j$ term which is $x_i$. Then \eqref{SexpIII} gives
\begin{equation}
S\le\tfrac12\sum_ix_i+\tfrac12\sum_i\Bigg[\sum_{j:j\ne
i}(\vvi)_{jj}-\Big(x_i+\sum_{j:j\ne i}(\vvi)_{jj}\Big)^2\Bigg]= \tfrac 12\sum_i[(x_i+y_i)-(x_i+y_i)^2]\,,
\label{SexpIV}
\end{equation}
where we denote
\eqn{
	y_i=\sum_{j:j\ne i}(\vvi)_{jj}\,.}[yi]
Finally introducing $z_i=x_i+y_i=\tr(\vvi)$, Eq.~\eqref{SexpIV} takes the form
\eqn{S\le\tfrac12\sum_i(z_i-z_i{\llnsp}^2)\,.}[SexpV]
Since $\max(z-z^2)=\frac14$, attained at $z=\frac12$, we finally obtain the
claimed inequality:
\eqn{S\le\tfrac18\lsp N\,.}[SBound]

The key idea in the above proof was to combine the second (positive) and the third (negative) terms in \SexpIII, which becomes possible after estimating the negative term as in  \keyIneq. If instead one were to neglect the negative term altogether, the resulting bound would have $C_N=O(N^2)$ as in the previous section, because the second positive term in \SexpIII contains $O(N^2)$ terms.

{\bf Remark.}
A simple modification of the above argument gives a bound on $S$ for
couplings whose beta-function is not zero, which may be of some interest. Namely, we have
\beq
S=\lambda_{ijkl}\lambda_{ijkl} \le \tfrac18\lsp N+ \calB\,,
\qquad \calB=\tfrac 13 \sum_i \beta_{iiii}+ \tfrac12\sum_{i\ne j} \beta_{iijj}\,,
\label{eq:BoundBeta}
\eeq
where $\beta_{ijkl}= -\lambda_{ijkl} +(\lambda_{ijmn}\lambda_{mnkl}+\text{2
permutations})$ is the rescaled beta-function.
In the proof, Eqs.~\bouI and \bouII get extra terms $\tfrac13 \beta_{iiii}$
and $\tfrac12 \beta_{iijj}$ in the right-hand side, which sum up to $\calB$.
The remaining estimates are unaffected.

\subsec{Improvements of the bound for \texorpdfstring{$N=1,2,3$}{N=1,2,3}}
\label{sec:N123}
The following small modification produces a slightly improved bound for $N=1,2,3$.
Note that in the above argument we treated the variables
$z_i$ entering the final estimate \SexpV as unconstrained, but in fact
$z_i=x_i+y_i$ where $x_i\in [0,\frac13]$ while $y_i$ can be bounded as
\eqn{|y_i|\le
\sqrt{N-1}\lsp\Big(\sum_{j\ne i}(\vvi)_{jj}{\llnsp}^2\Big)^{1/2}
=\sqrt{N-1}\lsp\sqrt{\tfrac13\lsp x_i-x_i{\llnsp}^2}\,,}[uppy]
where we have used the Cauchy--Schwarz inequality in the first step, and the second step follows from \bouI. So in fact \eqref{SexpIV} implies a more nuanced bound:
\eqn{S\le \tfrac12 N \kappa_N\,,}[imprBound]
where
\eqn{
\kappa_N=\max_{x\in[0,\frac13],y\in[0,\sqrt{N-1}(\frac13 x-x^2)^{1/2}]}
(x+y-(x+y)^2)\,.}
For $N\ge 4$ the maximum is attained at $z=x+y=\frac12$. We can e.g.\ take
$x=\frac14$, $y=\frac14$, and this satisfies the upper bound \uppy on $y$, so
$\kappa_N=\frac14$, and we go back to the original case of bound \SBound.
On the other hand, for $N=1,2,3$ we have
\eqn{
	z_N=\max_{x\in[0,\frac 13]} (x+ \sqrt{N-1}(\tfrac
	13x-x^2)^{1/2})<\tfrac12\,,}[]
and so
\eqn{\kappa_N = z_N-z_N^{\lsp 2}<\tfrac14\,,}[]
so that the new bound is stronger. We get
\eqn{\kappa_1=\tfrac 29,\qquad \kappa_2\approx 0.24047\,,\qquad
\kappa_3\approx 0.24801\,,}[]
where for $N=2,3$ extremization is carried out numerically.

The $N=1$ result is trivial and consistent with $S(\text{Ising})=\frac19$. The $N=2$ result is not particularly interesting because the fixed points are classified; see section \ref{sec:class}. Just as a sanity check, the two fixed points Ising$+$Ising and $O(2)$ both satisfy the bound, with $S(O(2))=\frac{6}{25}=0.24$ coming close to saturating it.

For $N=3$ the bound takes the form $S\le \frac 32 \kappa_3\approx 0.372015$ and is of some interest, since the full classification has
not yet been proven. Out of the known fixed points,
$S(O(3))=\frac{45}{121}\approx0.371901$ comes closest to saturating the
bound.

\subsec{Saturation of the bound for \texorpdfstring{$N\ge 4$}{N>4}}
\label{sec:sat}
In this section we will consider the case when the bound \SBound, or
equivalently \AStarBound, is best possible for $N\ge 4$. Turning this
around, we will try to understand if it is possible to find fixed points
that saturate \SBound.

Since the bound arose partly due to \keyIneq, to saturate it we
need to saturate \keyIneq, which happens if and only if the following sum
of off-diagonal terms vanishes:
\eqn{
	\sum_i (\vvi)_{jk} = \sum_i \lambda_{iijk}=0\,\qquad (j\ne k )\,.}[satI]
In addition, since $z_i=\tr(\vvi)$, and maximization of $z_i-z_i{\!}^2$
happens for $z_i=\tfrac12$, to saturate the bound we need
\eqn{\tr(\vvi)=\sum_j\lambda_{iijj}=\tfrac12\qquad\text{(no sum over
}i\text{)}\,.}[satII]
Eqs.~\satI and \satII can be summarized by saying that the bound will be saturated if and only if the fixed-point tensor satisfies
\eqn{\sum_i\lambda_{iijk}=\tfrac 12\lsp \delta_{jk}\,,}[satTrace]
a particular case with $z=\tfrac 12$ of the trace condition \eqref{satTrace1}. As explained in section \ref{sec:I42gen}, fixed points satisfying the trace condition are described by Eq.~\eqref{lamAns01} where $d_{ijkl}$ is a traceless symmetric tensor satisfying Eqs.~\eqref{ddpermsContr1}, \eqref{pqBound1}. Thus, fixed points saturating the bound are precisely solutions of \eqref{lamAns01}-\eqref{pqBound1} with $z=\tfrac 12$. This is true in full generality, e.g.~without assuming anything about the symmetry of the considered fixed points. As a check, notice that these equations imply
\eqn{\lambda_{ijkl}\lambda_{ijkl}= \tfrac{1}{2}N z (1-z)\,}[]
so that the bound is saturated if and only if $z=\tfrac 12$.

For $N=4$, substituting $z=\tfrac 12$ into \eqref{pqBound1} we get $p=q=0$. The obvious solution is $d_{ijkl}=0$. Hence the $O(4)$ fixed point saturates the bound~\cite{Osborn:2017ucf}. It also follows from \eqref{ddSq1} that it's the only solution.

Proceeding to $N\ge 5$, section \ref{sec:I42gen} also provided a way to construct examples of fixed points satisfying \eqref{lamAns01}-\eqref{pqBound1} using symmetry groups with $I_2=1, I_4=2$.
We would like to see when these examples saturate the bound. From \eqref{FPz1}, we conclude that
\beq
z=\tfrac 12\quad\Leftrightarrow \quad\rho=4(N-4)\,.
\eeq
Recall that by the general theory we have two real fixed points with $G$-symmetry if $\rho\ge 4(N-4)$, which coincide for $\rho=4(N-4)$. We thus see that the bound is saturated for groups with $I_2=1, I_4=2$ if and only if the two $G$-symmetric fixed points coincide. This equivalence is not an accident: it turns out that fixed points saturating our bound for $N\ne 4$, no matter their symmetry, necessarily have a marginal perturbation. The proof of this fact is postponed to Appendix \ref{sec:marginality}. For now let us go over the examples from section \ref{sec:I42ex} and recall when the coincidence happens.

We see that the cubic fixed points never saturate the bound. The tetrahedral fixed point saturates the bound if and only if $N=5$. The bifundamental fixed point saturates the bound for $m,n$ solving the Diophantine equation $R_{mn}=0$. There are infinitely many solutions given in Eq.~\eqref{Dioph}, the first one being $N=511=73\cdot 7$. The MN fixed point can only saturate the bound for $m=4$, when it factorizes into $O(4)$ fixed points, so that we don't get new fully interacting examples.

It would be interesting to look for more examples. We notice in this respect that a large number of $N=6$ fixed points satisfying \satTrace has been reported in~\cite{Hatch2}. The RG-stable fixed points found there have $z=\tfrac{6}{11}$ in our notation, and so they do not saturate the bound, although they come the closest, among the fixed points reported there, to doing so.

To summarize, our analysis implies that the bound $S\le \frac18\lsp N$ is
best possible for $N=4,5$, and for an infinite sequence of $N\ge 511$
obtained via \eqref{Dioph}. If one allows factorized fixed points, the
bound can also be trivially saturated putting together decoupled copies of
$O(4)$ and tetrahedral $N=5$ fixed points, i.e.~for all $N$ which can be
represented as a linear combination $4m+5n$ with nonnegative integers
$m,n$. This covers all integers $N\ge 4$ except $N=6,7,11$.\footnote{For
general natural numbers $a_1, a_2$ with $\gcd(a_1,a_2)=1$, the largest integer $N$ for which there is no representation $N=m_1 a_1+m_2 a_2$ with nonnegative integer $m_1, m_2$ is called the Frobenius number $g(a_1,a_2)$ of $a_1,a_2$. A theorem of Sylvester says that $g(a_1,a_2)=(a_1-1)(a_2-1)-1$ \cite{Coin}. In particular we have $g(4,5)=11$. Smaller numbers can be checked by hand.} For these values of $N$, Eqs.~\eqref{lamAns01}-\eqref{pqBound1} with $z=\tfrac 12$ can perhaps be investigated by brute force using computational algorithms of real algebraic geometry, although we have not attempted this.

\section{RG stability}
\label{sec:RGstab}

In this section we will present some general results about RG stability of
fixed points, mostly relying on the work of L.~Michel.

Let us remind the reader of some standard terminology. We call a fixed point
RG-stable if all quartic deformations around it are marginal or irrelevant.
This can be asserted by linearizing the beta-function equations,
\beq
\frac{d\lambda^I}{dt}=\beta^I(\lambda)\,,
\eeq
around the fixed point $\lambda_*$. RG stability means that the matrix $\Gamma^I{}_J=\del_J\beta^I$ has all eigenvalues $\gamma\ge0$ (we will see momentarily that all eigenvalues are real).

We will only study RG stability for the one-loop beta-function. Some deformations that are marginal at one loop may
become relevant or irrelevant at higher loop order. This phenomenon is  beyond the scope of our paper (see however section \ref{sec:zero} for some related comments).

RG stability should not be confused with potential stability discussed in section \ref{Stab}, where we showed that all fixed points have stable potential.

Clearly the trivial fixed point $\lambda=0$ is RG-unstable. Below we only examine nontrivial fixed points.

The $A$-function helps enormously to analyze RG-stability. Using Eq.~\Acubiccov we get
\beq
\del_J\beta^I = g^{IK} M_{KJ},\quad M_{KJ}=\del_K\del_J A\,,
\eeq
so the eigenvalue problem $\Gamma^I{\!}_J\lsp c^J=\gamma\lsp c^I$ for a generally 
nonsymmetric matrix $\Gamma$ is equivalent to the generalized symmetric eigenvalue problem
\beq
M_{IJ}\lsp c^J=\gamma\lsp g_{IJ}\lsp c^J
\eeq
for the Hessian matrix $M_{IJ}$ of $A$ at the fixed point. This has two
consequences. First, all eigenvalues $\gamma$ are real. Second, a fixed
point is RG-stable if and only if the Hessian evaluated at that fixed
point is positive semidefinite.

It will be interesting to inject an element of symmetry into the discussion of RG-stability. Let $H$ be a subgroup of $O(N)$ and consider the set of all quartic couplings $\Lambda_H$ which are $H$-invariant.
The set $\Lambda_H$ is a linear subspace of all coupling tensors, and it is preserved by RG evolution.

Suppose $\lambda_*\in \Lambda_H$ is a fixed point. Notice that the symmetry
group of $\lambda_*$ is at least as large as $H$ but may be strictly
larger. It is interesting\footnote{This is also physically important since
the set of allowed perturbations of the microscopic Hamiltonian is often
restricted by symmetry.} to consider RG-stability of $\lambda_*$ with
respect to perturbations belonging to $\Lambda_H$, which we will call
RG-stability \emph{within} $\Lambda_H$. By the same argument as above, this
property holds if and only if $A$ restricted to $\Lambda_H$ has positive
semidefinite Hessian at $\lambda_*$. The unrestricted RG stability
corresponds to $H=\{ 1\}$, $\Lambda_H=\{\text{all couplings}\}$.

\subsection{Uniqueness of RG-stable fixed point}

{\bf Theorem.} (Michel \cite{Michel:1983in}) Suppose $\lambda_1,\lambda_2\in \Lambda_H$ are two nontrivial nonidentical fixed points. Consider the value of the $A$-function at them: $A(\lambda_1)$, $A(\lambda_2)$. Then
\begin{itemize}
	\item if $A(\lambda_1)\ne A(\lambda_2)$, then the fixed point with larger $A(\lambda)$ is RG-unstable within $\Lambda_H$ (while the other one may or may not be RG stable),
	\item if $A(\lambda_1)= A(\lambda_2)$, then both fixed points are RG-unstable within $\Lambda_H$.
\end{itemize}
As a consequence, there is at most one fixed point RG-stable within
$\Lambda_H$.\footnote{We stress that, as almost all results in this paper,
this theorem is valid at one loop. Extra RG stable fixed points may appear in higher orders of the $\varepsilon$-expansion, or when this expansion is extrapolated to $\vareps=1$. See \cite{Vicari:2006xr} for a discussion.}

{\bf Remark.} ``Nonidentical" in the statement of the theorem means simply that $\lambda_1$ and $\lambda_2$ are unequal tensors. Nonidentical fixed points may be physically equivalent if they are related by an $O(N)$ transformation. The theorem still applies in this case. This remark will be important in the next section.

\emph{Proof.} We give a pedagogical version of the original proof
  in~\cite{Michel:1983in}; another presentation can be found
in~\cite{Vicari:2006xr}, but it does not cover the case $A(\lambda_1)=
A(\lambda_2)$, which is important for applications in section \ref{sec:criteria}.

The main idea is to consider the restriction of $A$ to the two-plane within $\Lambda_H$ spanned by $\lambda_1$ and $\lambda_2$. The Hessian of restricted $A$ is evaluated by explicit computation, and the statements of the theorem follow.

To avoid getting lost in indices, let us denote for any symmetric
four-tensors $u,v,w$,
\begin{gather}
(u,v)=u_{ijkl}v_{ijkl}\,,\\
(u,v,w)= u_{ijkl}v_{klmn} w_{mnij}\,.
\end{gather}
Notice that $(u,v)$ and $(u,v,w)$ do not depend on the order of the arguments.
Contracting the beta-function equations expressing the fact that $\lambda_1,\lambda_2$ are fixed points,
we get the following auxiliary results:
\begin{subequations}
\begin{gather}
(\lambda_i,\lambda_i,\lambda_i)=\tfrac 13\vareps (\lambda_i,\lambda_i)\,,\qquad i=1,2\,,\\
(\lambda_1,\lambda_1,\lambda_2)=(\lambda_2,\lambda_2,\lambda_1) = \tfrac 13\vareps (\lambda_1,\lambda_2)\,.
\end{gather}
\label{eq:auxMichel}
\end{subequations}
The first of these equations is \betaContrLam, and the second one is a simple generalization.

 In the above notation
 \beq
 A(\lambda)=-\tfrac 12 \veps \lsp(\lambda,\lambda)+(\lambda,\lambda,\lambda)\,.
 \eeq
 Using (\ref{eq:auxMichel}a), we recover \AStar
 \beq
 A(\lambda_i)=-\tfrac 16\veps\lsp (\lambda_i,\lambda_i)\,,\qquad i=1,2\,.
 \eeq
 We will assume without loss of generality that $(\lambda_2,\lambda_2)\ge (\lambda_1,\lambda_1)$ and will show that $\lambda_1$ is unstable.

 We are interested in $A$ restricted to the two-plane spanned by $\lambda_1$ and $\lambda_2$, parametrized as
 \beq
A(\lambda_1+s\lambda_1+t\lambda_2)\,.
 \eeq
This is a cubic polynomial in $s,t$ and using \reef{eq:auxMichel} we could evaluate all coefficients.
For our purposes of extracting the Hessian, we just evaluate the part
quadratic in $s,t$, which comes out equal to
\beq
\tfrac 12\veps\lsp (\lambda_1,\lambda_1) s^2 +\veps\lsp (\lambda_1,\lambda_2)
st+\veps\lsp\big[ (\lambda_1,\lambda_2)-\tfrac 12
(\lambda_2,\lambda_2)\big]t^2\,,
\label{eq:quadpart}
\eeq
corresponding to the Hessian matrix
\beq
M=\veps \begin{pmatrix}
a_1 & b\\
b & 2b- a_2
\end{pmatrix}\,,\qquad a_i=(\lambda_i,\lambda_i),\  b=(\lambda_1,\lambda_2)\,.
\eeq
We would like to show that this Hessian is not positive semidefinite. Since the matrix element $a_1>0$, one of the eigenvalues is positive and we need to show that the second is negative, which will be the case if and and only if the determinant is negative. We have
\beq
\det M/\veps^2= 2 ba_1-a_1 a_2-b^2=-(a_1-b)^2-a_1(a_2-a_1)\,.
\eeq
If $a_2>a_1$, this is negative.\footnote{In this case the quadratic part
\reef{eq:quadpart} is negative along the line $t=-s$, moving from
$\lambda_1$ in the direction of $\lambda_2$ \cite{Michel:1983in}.} If
$a_2=a_1$, this is negative unless $b=a_1$, however the latter is
impossible,
since $(\lambda_1,\lambda_1)=(\lambda_2,\lambda_2)=(\lambda_1,\lambda_2)$ implies $(\lambda_1-\lambda_2,
\lambda_1-\lambda_2)=0$ and we are assuming $\lambda_1\ne \lambda_2$. So in all cases $\det M$ is negative. This completes the proof that $\lambda_1$ is unstable.

\subsection{Criteria for RG instability}
\label{sec:criteria}

To apply Michel's theorem, we need to have two fixed points. But suppose we are given only one fixed point $\lambda_* \in \Lambda_H$. We can still sometimes use the theorem to conclude that $\lambda_*$ is unstable, using an idea of \cite{MichelToledano}. Consider the orbit of $\lambda_*$ under the action of $O(N)$, denoted $O(N)\cdot \lambda_*$.

{\bf Fact 1.} If this orbit intersects $\Lambda_H$ at some other point
besides $\lambda_*$, then the fixed point $\lambda_*$ is RG-unstable within $\Lambda_H$.

\begin{figure}[t!]
	\centering
	\includegraphics[width=0.5\linewidth]{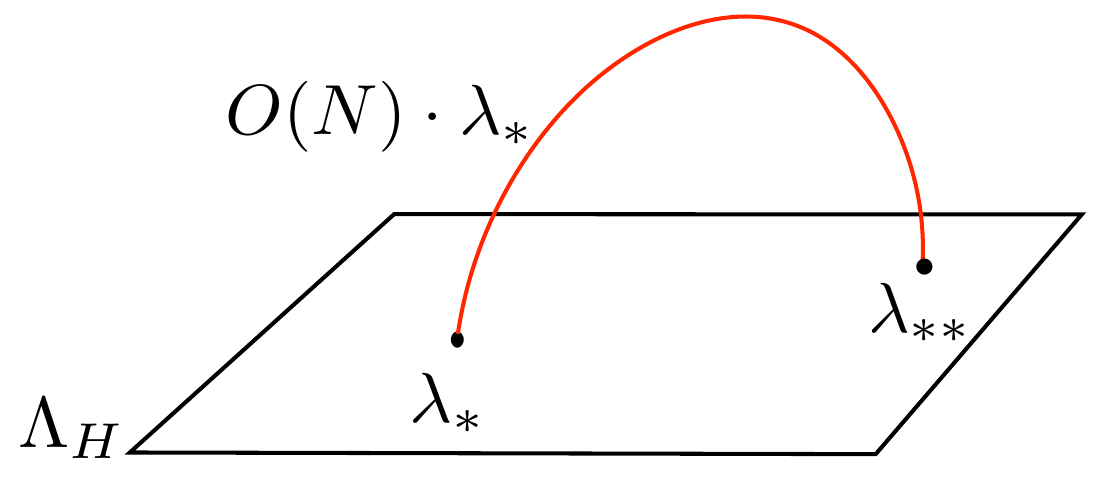}
	\caption{\label{fig:orbit} If the orbit $O(N)\cdot \lambda_*$ has any other intersections with $\Lambda_H$ apart from $\lambda_*$, as in this figure, the fixed point $\lambda_*$ cannot be RG-stable within $\Lambda_H$.}
\end{figure}

\emph{Proof.} (see Fig.~\ref{fig:orbit}) Let $\lambda_{**}\ne \lambda_*$ be
another intersection. Since $\lambda_{**}$ is obtained from $\lambda_*$ by
an $O(N)$ transformation, they have the same $A$-function:  $A(\lambda_*)=
A(\lambda_{**})$. By Michel's theorem, $\lambda_*$ is then RG-unstable. Of
course the two fixed points are completely physically equivalent. However,
$\lambda_*$ and $\lambda_{**}$ are two different \emph{tensors}, and so Michel's theorem is applicable. QED

We thus have a sufficient condition for an RG fixed point to be unstable. As we will see now, applicability of this condition depends just on $H$ and on the symmetry group of $\lambda_*$ which we denote $G_*$. Notice that $H\subset G_*\subset O(N)$, but that $G_*$ may be strictly larger than $H$. Suppose we have
\begin{gather}
\lambda_{**}=g_0\cdot \lambda_*\in \Lambda_H, \quad
\lambda_{**}\ne \lambda_*\,,
\end{gather}
where $g_0\cdot$ denotes the group action of an element $g_0\in O(N)$ on the tensor. The condition $\lambda_{**}\ne \lambda_*$ is equivalent to $g_0\notin G_*$.
On the other hand the condition $\lambda_{**}\in \Lambda_H$ is equivalent to
\beq
h\cdot (g_0\cdot \lambda_*) = g_0\cdot \lambda_* \text{ for any } h\in H
\eeq
or equivalently, multiplying both sides by $g_0^{-1}$,
\beq
(g_0^{-1}h g_0)\cdot \lambda_* = \lambda_* \text{ for any } h\in H\,.
\eeq
The latter condition can be expressed as
\beq
g_0 ^{-1} H g_0 \subset G_*\,.
\eeq
We thus have an equivalent formulation of Fact 1:

{\bf Fact 2.} Let $H\subset G_*$ be two subgroups of $O(N)$. Suppose there exists an $O(N)$ element $g_0$, such that $g_0\notin G_*$ and $g_0 ^{-1} H g_0 \subset G_*$. Then any fixed point $\lambda_*\in \Lambda_H$ with symmetry $G_*$ is RG-unstable within $\Lambda_H$.

Consider now some simpler but strictly weaker conditions. Recall that the \emph{normalizer} of any subgroup $H$ of $O(N)$ is defined as
\beq
N(H)=\{g:g ^{-1} H g\subset H\}\,.
\eeq
The normalizer is itself a subgroup of $O(N)$. Clearly $N(H)\supset H$ but it may be strictly larger.

Then we have

{\bf Fact 3.} Suppose that $H\subset G_*$, and that the normalizer $N(H)$ is strictly larger than $G_*$. Then any fixed point $\lambda_*\in \Lambda_H$ with symmetry $G_*$ is RG-unstable within $\Lambda_{H}$.

\emph{Proof.} This is strictly weaker than Fact 2. Take $g_0\notin G$, $g_0\in N(H)$. The latter implies by definition $g_0 ^{-1} H g_0 \subset H$ (and thus $\subset G_*$).

Specializing Fact 3 to $H=G_*$ we get:

 {\bf Fact 4.} \cite{MichelToledano} Suppose that the normalizer $N(G_*)$ is strictly larger than $G_*$. Then any fixed point $\lambda_*$ with symmetry $G_*$ is RG-unstable within $\Lambda_{G_*}$ (and thus within $\Lambda_{H}$ for any $H\subset G_*$).

 These criteria have many applications, some of which have been explored in \cite{MichelToledano,TMTB}.
 Here we will consider only one application. Consider unrestricted RG stability: $H=\{1\}$, $\Lambda_H=\{\text{all couplings}\}$.
By Fact 1, a fixed point $\lambda_*$ may be unrestricted RG stable only if its entire $O(N)$ orbit consists of just one point, which means that $\lambda_*$ is $O(N)$ invariant. [We can also see this from Fact 3 since $N(H)=O(N)$.]

Of course there is only one nontrivial fixed point with $O(N)$ symmetry---the $O(N)$ fixed point.
It is known to be one-loop stable
 for $N=2,3,4$,\footnote{For $N=3,4$ the cubic deformation is marginal and at higher orders it becomes irrelevant for $N=3$ and relevant at $N=4$. Recall that higher order stability is beyond our scope in this paper.}  while for $N>4$ it is unstable as it flows e.g.\ to the cubic fixed point for these
 $N$. So one
 consequence is that for $N> 4$ there are no unrestricted RG-stable fixed points.

\subsec{Zero RG eigenvalues: divergences of broken currents vs marginal operators}
\label{sec:zero}
Here we would like to discuss and resolve a potential confusion related to the interpretation of zero eigenvalues of the linearized beta-function equation.

Consider a fixed point $\lambda_*$ with symmetry group $G_*\subset O(N)$.  Let $G_{conn}\subset SO(N)$ be the connected component of $G_*$ containing the unity. For the free and the $O(N)$ fixed points, and only for these two, we have $G_{conn}=SO(N)$. For any other fixed point $G_{conn}$ is strictly smaller than $SO(N)$, and this is the case we wish to examine.

As usual we choose a basis of $SO(N)$ generators as $\{V_k, B_l\}$ where $V_k$ are generators of $G_{conn}$, called unbroken, while the `broken generators' $B_l$ are a remaining set of generators completing the basis.
The number of broken generators $N_B=\dim (SO(N))-\dim(G_{conn})>0$ by assumption.
Acting on the fixed point $\lambda_*$ with broken generators we generate a manifold $\calM$ of tensors of dimension $N_B$. By covariance of the beta-function equation, all tensors of $\calM$ have zero beta-function. These are all RG fixed points, physically equivalent to $\lambda_*$ but described by different tensors.

Consider now a perturbation of $\lambda_*$ by a tensor $\delta\lambda$, in a direction tangent to $\calM$. By the above discussion, any such perturbation will be an eigenperturbation of the linearized RG equation, with eigenvalue zero.
Notice that this statement will be true to any order in perturbation
theory. Does this mean we should think of such a perturbation as an exactly
marginal operator? The answer is negative---these zero eigenvalues have a different interpretation.

To understand what's going on, we should consider the fate of current operators. The multiscalar theory we are studying,
\beq
\tfrac 12 \del^\mu \phi_i\lsp\partial_\mu\phi_i +\tfrac 1{4!} \lambda_{ijkl} \phi_i\phi_j\phi_k\phi_l
\eeq
has current operators
\beq
J_\mu = \omega_{[ij]}\lsp\phi_i \del_\mu \phi_j\,,
\eeq
where $\omega_{[ij]}$ parametrizes the $SO(N)$ algebra. Using the equation of motion
\beq
\del^2\phi_i =\tfrac1{3!} \lambda_{ijkl} \lsp\phi_j\phi_k\phi_l\,,
\eeq
the conservation equation for the current takes the form
\begin{gather}
\del^\mu J_\mu = \calO_{\delta\lambda} =
(\delta\lambda)_{ijkl}\lsp\phi_i\phi_j\phi_k\phi_l\,,\qquad (\delta\lambda)_{ijkl} = \omega_{im}\lambda_{mjkl}+\text{3 permutations.}
\end{gather}
At the fixed point, the currents corresponding to the unbroken generators will have $\delta\lambda=0$ and will be conserved.
On the other hand the broken generators are those for which $\delta\lambda\ne 0$. Currents corresponding to the broken generators are not conserved, and their divergence is given the quartic operators $\calO_{\delta\lambda}$. The corresponding $\delta\lambda$'s are precisely the deformations along the manifold $\calM$ discussed above.\footnote{\label{note:traceless}Notice that any such $\delta\lambda$ satisfies the `double tracelessness' condition $(\delta\lambda)_{iijj}=0$. This is because the double trace is invariant under an infinitesimal $SO(N)$ transformation. This remark will be useful in Appendix \ref{sec:marginality}.} Since $\calO_{\delta\lambda}$ is a total derivative operator, perturbing by $\int \calO_{\delta\lambda}$ leaves the theory unchanged. This is not the same as perturbing by an exactly marginal operator, which leads from one CFT to another, strictly different CFT.

In light of the above, it would not be even correct to ask if  $\int \calO_{\delta\lambda}$ is an irrelevant, relevant, or marginal perturbation, since it's not a perturbation at all! It still makes sense to ask what is the scaling dimension of $\calO_{\delta\lambda}$ (as determined e.g.~from the two point function),
but linearized RG teaches us nothing in this respect. Indeed, the RG eigenvalue being zero means that if we add
\beq
g \int d^{\llsp 4-\veps}x\, \calO_{\delta\lambda}(x)
\label{add}
\eeq
to the action and perform an RG step, the coefficient $g$ does not change.
However, since \reef{add} is identically zero, adding it to the action achieves strictly nothing. That $g$ does not change contains no nontrivial information and cannot
be used to draw conclusions about the scaling dimension of
$\calO_{\delta\lambda}$. To study this scaling dimension using the RG, one would need more nuanced probes, for example
adding the term like \reef{add} but with a space-dependent coupling $g(x)$;
see below.

\begin{figure}[t!]
  \centering
  \includegraphics[width=0.5\linewidth]{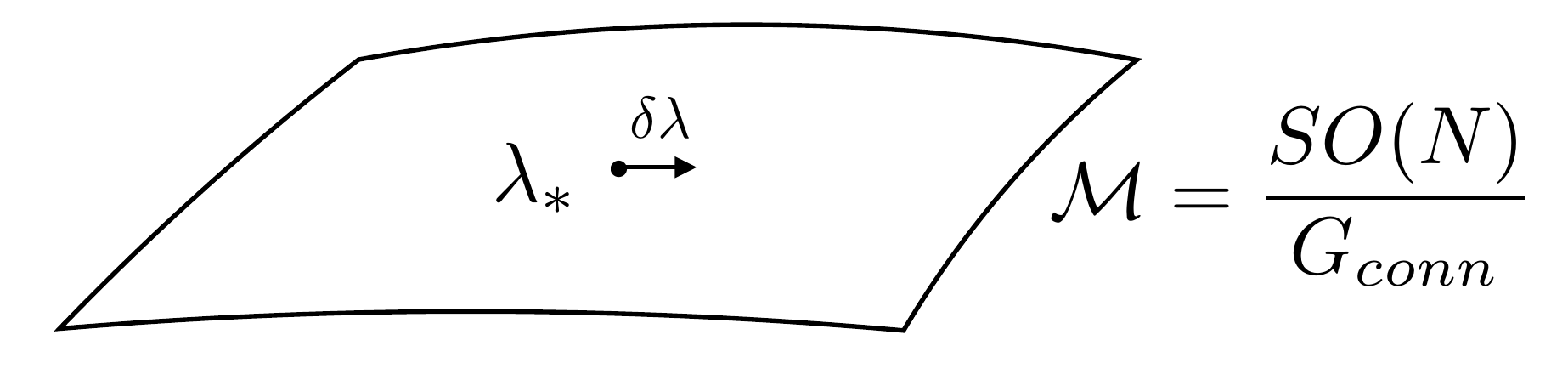}
  \caption{\label{fig:broken} For fixed point perturbations by quartic
  scalar interactions corresponding to broken symmetry generators,
linearized RG naively predicts that they are marginal. Instead such
deformations correspond to total derivative operators of scaling dimension
larger than $d$; see the text.}
\end{figure}

Instead, a general conclusion about the scaling dimensions of the
$\calO_{\delta\lambda}$ operators can be made by relating them to the
broken currents $J_\mu$.  Being broken, these currents will pick up
anomalous dimensions $\gamma_J$. As usual for currents, this will first
happen at two loops in the $\veps$ expansion,
$\gamma_J=\text{O}(\vareps^2)$. Importantly, these anomalous dimensions
will be positive $\gamma_J>0$ as a consequence of unitarity.\footnote{As
  recently discussed in \cite{Hogervorst:2015akt} the theory in $4-\veps$
  dimensions is not quite unitary. This absence of unitarity, however,
  affects only the high-dimension sector of the theory, while at low
dimensions unitarity constraints still apply.} Their divergences
$\calO_{\delta\lambda}$ will therefore have dimensions $d+\gamma_J$ at the
IR fixed point. Based on their dimension, one could say that these operators
are irrelevant, but as mentioned above the term `irrelevant' is not
applicable to total derivative operators. See Fig.~\ref{fig:broken}.

To avoid any misunderstanding, we confirm that there is no subtlety for all the other zero eigenvalues of the linearized RG evolution, those whose eigenvectors cannot be obtained by acting on $\lambda_*$ with an $SO(N)$ generator. Their eigenvectors correspond to perturbations marginal in the one-loop approximations, which may become relevant or irrelevant in higher-loop approximation.

Let us illustrate the above discussion with a concrete example for $N=2$. As mentioned in section \ref{sec:class}, in this case there is a complete classification of fixed points \cite{Osborn:2017ucf}. We have
the $O(2)$ fixed point, the free fixed point, the direct product of the Ising fixed point and a
free theory, and the direct product of two Ising fixed points.

Let us focus on the latter case, which obviously does not have $O(2)$
symmetry. Let us study this theory in the frame where $\phi_1$ and $\phi_2$
are the two Ising fixed point fields (i.e.~$\lambda_{1111}=\lambda_{2222}$
are the only two nonzero components of the fixed point coupling tensor). We
can consider arbitrary linearized perturbations around this fixed point.
The space of $N=2$ symmetric four-tensors being five-dimensional,
we have $5\times 5$ stability matrix. It has a zero eigenvalue, corresponding to the operator
$\calO_- = \phi_1\phi_2{\!}^3-\phi_1{\!}^3\phi_2$~\cite{Osborn:2017ucf}. This operator is precisely the result of acting on $\lambda_*$ with an infinitesimal rotation. Using the
equations of motion we easily check that this is a descendant:
\eqn{\calO_-= \phi_1\phi_2{\!}^3-\phi_1{\!}^3\phi_2\sim\partial_\mu(\phi_1
	\partial^\mu\phi_2 - \phi_2\lsp\partial^\mu\phi_1)\,.}[primDesc]
Since $\calO_-$ is a total derivative, a corresponding zero RG eigenvalue
does not imply that its dimension is $d$.
Given that we are considering a factorized theory, the dimension of the primary
$\phi_1\lsp\partial_\mu\phi_2-\phi_2\lsp\partial_\mu\phi_1$ is computed immediately as
$d-1+\gamma_{\phi_1}+\gamma_{\phi_2}$, hence the dimension of $\calO_-$ is $d+\gamma_{\phi_1} +\gamma_{\phi_2}$. This is also consistent with the fact that
the dimension of $\calO_-$ should be equal, in the factorized theory, to
that of $\calO_+=\phi_1\phi_2{\!}^3
+\phi_1{\!}^3\phi_2$.
Notice that since the operator $\calO_+$ is not a total derivative its
dimension is correctly predicted by linearized RG methods (it's the
second eigenvector in \cite[Eq.~(3.12)]{Osborn:2017ucf}).

It is interesting to extract the dimension of $\calO_-$ in a more
direct way.  As mentioned this can be done by performing a deformation of
our theory with this total-derivative operator but with an $x$-dependent
coupling. In that case, we can no longer integrate by parts to remove the
deformation. Renormalization with space-dependent couplings requires new
counterterms as explained in~\cite{Jack:1990eb,
	Fortin:2012hn}.  The
	operator
$J^\mu=\omega_{ij}J^\mu_{ij}=\omega_{ij}\llsp\phi_i\llsp\partial^\mu\phi_j$,
$\omega_{ij}=-\omega_{ji}$ has dimension $d+\gamma_{ij}$, where the
anomalous dimension is given by
\eqn{\gamma_{ij}=(\rho_{klmn})_{ij}(\omega\lambda)_{klmn}\,,\qquad
	(\omega\lambda)_{ijkl}=\omega_{im}\lambda_{mjkl}+\text{permutations}\,,
}[gamEq]
with
\eqn{(\rho_{klmn})_{ij}=(N^1_{klmn})_{ij}
	+\lambda_{pqrs}\frac{\partial}{\partial\lambda_{pqrs}}(N^1_{klmn})_{ij}\,,
}[rhoEq]
where $(N^1_{klmn})_{ij}$ is the $1/\veps$ pole of the counterterm
$(N_{klmn})_{ij}\lsp\partial^\mu\lambda_{klmn}\llsp
\phi_i\llsp\partial_\mu\phi_j$ required in the theory with space-dependent
couplings. This receives contributions order by order in perturbation
theory, and it has been computed that at two loops~\cite{Jack:1990eb,
Fortin:2012hn}
\eqn{(N^1_{klmn})_{ij}=-\tfrac{1}{24}
	(\lambda_{iklm}\delta_{jn}-\lambda_{jklm}\delta_{in})\,.}[]
From this and \rhoEq we find for \gamEq, at two loops,
\eqn{\gamma_{ij}=-\tfrac{1}{12}(\lambda_{iklm}\lambda_{klmn}
	\llsp\omega_{jn}-\lambda_{jklm}\lambda_{klmn}\llsp
	\omega_{in})-\tfrac12\llsp\lambda_{ikmn}\lambda_{jlmn}\llsp\omega_{kl}\,.
}[gamTwoLoops]
In our case of two fields $\phi_1$ and $\phi_2$, where
$\omega_{ij}=\epsilon_{ij}$, \gamTwoLoops gives, at the decoupled Ising
fixed point,
\eqn{\gamma_{12}=\tfrac{1}{12}(\lambda_{1111}^2
	+\lambda_{2222}^2)\,,}[anDimJ]
exactly as expected in order to give the dimension of $\calO_-$ as
$d+\gamma_{\phi_1}+ \gamma_{\phi_2}$. Note that although we have used the
fact that $\calO_- \sim\partial_\mu(\phi_1\partial^\mu\phi_2 -
\phi_2\lsp\partial^\mu\phi_1)$, the result \anDimJ arose directly from the
general expression \gamTwoLoops, i.e.\ without using our knowledge of
$\gamma_{\phi_1}$ and $\gamma_{\phi_2}$ in the decoupled Ising theory.

\newsec{Conclusion}
A very important characteristic of low loop order beta-function expressions
obtained in scalar theories in $4-\veps$ dimensions is that they arise from
a gradient, i.e.\
\eqn{\beta^I=g^{IJ}\partial_J A\,.}[]
The main result of this paper is a general bound on the critical value of
$A$ at leading loop order, given by
\eqn{A_\ast\ge-\tfrac{1}{12}N\lsp\varkappa_N\lsp\veps^3\,,}[AboundConc]
where $N$ is the length of the vector order parameter $\phi_i$ and
\eqn{\varkappa_N=\begin{cases}
    \tfrac29 & N=1\,,\\
    0.24047 & N=2\,,\\
    0.24801 & N=3\,,\\
    \tfrac14 & N\ge4\,.
\end{cases}}[]
This bound demonstrates that although RG flows toward the IR cause $A$ to
decrease, this cannot continue indefinitely for flows leading to a fixed
point. More specifically, CFTs that are closest to saturating or actually
saturate the bound \AboundConc cannot be deformed by relevant operators and
flow to other CFTs. Such deformations, if they exist, can give rise only to
flows running away to large couplings and/or unstable potentials. A
physical interpretation of such runaway flows is a first-order phase
transition.

A perhaps more desirable way to phrase our bound would be to express it in
terms of a physical quantity.  The coefficient of the stress-energy tensor
two-point function, $C_T$, provides us with a good candidate. At leading
order there is a general result~\cite{Osborn:2017ucf},
\eqn{\frac{C_T}{C_{T,\llsp\text{scalar}}}=N-\tfrac{5}{36}\lambda_{ijkl}
\lambda_{ijkl}\,,}[]
where $C_{T,\llsp\text{scalar}}$ is the result for a single free scalar.
Our bound can then be cast in the form
\eqn{\frac{C_T}{C_{T,\llsp\text{scalar}}}\ge N\left (1- \tfrac{5}{72}\lsp\varkappa_N\lsp\veps^2\right)\,.}[]

We have found that certain theories saturate the bound \AboundConc.
Saturation of the bound can be achieved when fixed points with
the same global symmetry that move in coupling space as $N$ is varied
coincide. The $N=1,4$ cases are special---the bound is then saturated by
the Ising and $O(4)$ models respectively. For $N=2$ the bound cannot be
saturated. For $N=3$ we do not know of a theory that saturates the bound,
as is also the case for theories with $N=6,7,11$. For $N=5$ the tetrahedral
theory saturates the bound. Further nontrivial examples of bound saturation
arise by the bifundamental fixed point, e.g.\ for $N=511$. It would
be interesting to compile a complete list of theories that can saturate our
bound. In general, fixed points saturating the bound at $N\ne 4$ have a
marginal deformation (see appendix \ref{sec:marginality}).

It is obviously of interest to extend our results in other directions.
Within the $\veps$ expansion, one could examine the fate of the bound when
fermions are added. Even when Yukawa couplings are considered, it is still
the case that the flow is gradient at leading order~\cite{Wallace:1974dy}.
When applied to the results of~\cite{Fei:2015oha} our bound shows that
$\tilde{F}_{\text{UV}}-\tilde{F}_{\text{IR}}$ is bounded from above for
scalar fixed points. It would be interesting to find a physical argument to
justify this upper bound, and examine possible generalizations using the
methods and results of~\cite{Fei:2015oha}.

It is also important to examine the fate of the bound beyond leading order.
We remind the reader that the RG flow is gradient even at two loops in a
theory with scalars and fermions in $4-\veps$ dimensions~\cite{Jack:1990eb,
Fortin:2012ic}. In a theory with only scalars at two loops we have
\eqn{A=-\tfrac12\lsp\veps\lsp\lambda_{ijkl}\lambda_{ijkl}
+\lambda_{ijkl}\lambda_{klmn}\lambda_{mnij}
+\tfrac{1}{12}\lambda_{ijkl}\lambda_{jklm}\lambda_{mnpq\vphantom{l}}
\lambda_{npqi\vphantom{l}}
-\tfrac32\lambda_{ijkl}\lambda_{kmnp}\lambda_{lmnq}\lambda_{pqij}\,.}[]
Using the beta-function equation and the expansion
$\lambda_{ijkl}=a_{ijkl}\lsp\veps
+b_{ijkl}\lsp\veps^2+\text{O}(\veps^3)$ we find, at a fixed point,
\eqn{A_\ast=-\tfrac16\lsp\veps^3\lsp a_{ijkl}
a_{ijkl}
+\lsp\veps^4\big(\tfrac1{12}\lsp
a_{ijkl}a_{jklm}a_{mnpq\vphantom{l}}
a_{npqi\vphantom{l}}
-
\tfrac{3}{2}\lsp a_{ijkl}a_{kmnp}
a_{lmnq} a_{pqij}
\big)+\text{O}(\veps^5)\,,}[]
extending the result \AStar beyond leading order. It is interesting that
using the one-loop beta-function equation we were able to eliminate
$b_{ijkl}$ from the $\veps^4$ term. However, a bound
pertaining to the $\veps^4$ correction is not obvious. We hope to explore
this possibility in future work.

In Sec.~\ref{Review} we provided a review of some known fixed points in
$d=4-\veps$. For specific choices of $N$ there are many more fixed points
one encounters, see e.g.\ \cite{TMTB} for $N=4$ and \cite{Hatch, Hatch2}
for $N=6$. The study of these fixed points in $d=4-\veps$ but also in $d=3$
with the conformal bootstrap~\cite{Poland:2018epd} is of obvious interest
and importance. As of this writing there have been only a couple of
attempts in this direction~\cite{Rong:2017cow, Stergiou:2018gjj}.

While the $\veps$ expansion of scalar theories around four dimensions has
had a long history of active research, general statements about fixed
points that can be obtained within it are rather scarce.  In this work we
proved the bound~\AboundConc, and discussed a few other general statements,
some of which have appeared in the work of L.~Michel.  We also provided a
quick review of some famous scalar fixed points. We hope that our work will
provide at least an $\veps$ step towards the goal of fully classifying
scalar CFTs in $d=4-\veps$ dimensions.
\newpage

\ack{We would like to thank H.~Osborn for illuminating discussions and
comments. We would also like to thank S.~Giombi, J.~Gracey, M.~Hogervorst,
I.~Klebanov, H.~Osborn and E.~Vicari for comments on the manuscript. AS
would like to thank IHES and SR for hospitality. SR is supported by the
Simons Foundation grant 488655 (Simons Collaboration on the Nonperturbative
Bootstrap), and by Mitsubishi Heavy Industries as an ENS-MHI Chair holder.}
SR would like to thank the Isaac Newton Institute for Mathematical Sciences
for support and hospitality during the programme ``Scaling limits, rough
paths, quantum field theory'' when work on this paper was undertaken. This
work was supported by: EPSRC grant number EP/R014604/1.

\begin{appendices}
\section{Saturation of the bound and marginality}
\label{sec:marginality}

It was observed in section \ref{sec:sat} that saturation of the bound on $A$ seems to go hand in hand with pairs of fixed points colliding.
We saw these collisions in families of fixed points having fixed symmetry,
but one may wonder if there is more general significance to this
observation. The following result gives an affirmative answer:

{\bf Fact.} For $N\ne 4$, a fixed point saturating the bound necessarily has a marginal deformation (independently of any symmetry assumptions).
\\[-10pt]

The connection to fixed point collisions is obvious, since colliding fixed
points are well-known to have marginal deformations. Intuitively we can
understand this by using the following toy model. Suppose we have a family
of RG flows continuously depending on a parameter $y$, and for each $y<y_0$
there are two fixed points that collide for $y=y_0$. Generically, close to
the collision point we can focus on just one coupling $g$ whose running is described by the phenomenological beta-function
\beq
\beta(g)= y-y_0 +(g-g_0)^2\,.
\eeq
The precise value of $g_0$ is not important. What is important is that for
$y<y_0$ we have two fixed points at $g=g_0\pm \sqrt{y_0-y}$ and the
dimension of the operator which couples to $g$ is $\Delta=d+\beta'(g)=d\pm
2\sqrt{y_0-y}$ at each of them. For $y=y_0$, when fixed points collide, this operator is marginal.\footnote{See \cite{Gorbenko:2018ncu} for a more detailed review, and for what happens at $y>y_0$ when fixed points go to the complex plane.}

Let us now move past this toy model and prove the above fact. We use the
general characterization of fixed points saturating the bound given in
section \ref{sec:sat}. These are precisely fixed points that can be written
in the form \reef{lamAns01}-\eqref{pqBound1} with a symmetric traceless
tensor $d_{ijkl}$ and $z=\tfrac 12$. For convenience we copy these
conditions here:
\eqn{\begin{gathered}
\lambda_{ijkl}=\tfrac{1}{N+2}z \lsp T_{ijkl}+d_{ijkl}\,,\qquad
	T_{ijkl}=\delta_{ij}\delta_{kl}+\delta_{ik}\delta_{jl}
	+\delta_{il}\delta_{jk}\,,\\
 d\vee d = \tfrac 13(p\lsp T + q\lsp d) \,,\qquad
p=\tfrac{1}{N+2}\lsp z\big(1-\tfrac{N+8}{N+2}\lsp z\big)\,,\qquad
q= 1 - \tfrac{12}{N+2}\lsp z\,.
\end{gathered}}[dveed]
Here, following \cite{Michel:1983in}, we introduced the vee product of two symmetric four-tensors $u,v$ which is the symmetric four tensor $u\vee v$ defined by
\beq
(u\vee v)_{ijkl}= \tfrac 16(u_{ijmn}v_{mnkl} + u_{ikmn}v_{mnjl}+u_{ilmn}v_{mnjk}+u\leftrightarrow v)\,.
\eeq
We are interested in $N\ne 4$ because for $N=4$ the $O(N)$ fixed point ($d_{ijkl}=0$) is the only solution.

We would like to find a perturbation $u$ of the fixed point that is marginal. By section \ref{sec:RGstab}, a marginal direction is a zero eigenvector of the Hessian $H$ of the $A$-function around the fixed point, which means
\beq
H_{uv}=0\quad\text{for any}\ v\,.
\eeq
It will be convenient to evaluate this matrix element in an index-free way, as
\beq
H_{uv}=\del_s\del_t A(\lambda+t u+ sv)\,.
\eeq
Expressing the $A$-function in the notation of section \ref{sec:RGstab} as (we set $\vareps=1$)
\beq
A=-\tfrac 12 (\lambda,\lambda)+(\lambda,\lambda,\lambda)\,,
\eeq
we find
\beq
H_{uv}=-(u,v)+6(\lambda,u,v)=(-u+6\lsp\lsp\lambda\vee u,v)\,.
\eeq
This vanishes for any $v$ if and only if
\beq
\lambda\vee u = \tfrac 16 u\,,
\eeq
which is therefore the condition for the existence of a marginal direction $u$.

Let us show that this equation has a solution. We will look for a solution in the form
\beq
u=T+ x\lsp d
\eeq
for some unknown $x$. It's easy to compute
\beq
T\vee T=\tfrac {N+8}3\lsp T\,,\qquad T\vee d = 2\lsp d\,,
\eeq
while $d\vee d$ is given in \dveed. The parameter $x$ has to satisfy two
linear equations. For $z\ne \tfrac 12$ one finds that there is no solution,
while precisely for $z=\tfrac 12$ the two equations become linearly dependent and one finds
\beq
x=-\tfrac{12(N+2)}{N-4}\,.
\eeq
Therefore, we have a marginal direction as claimed. Notice that the worries from section \ref{sec:zero} do not apply: $u$ cannot be obtained from $\lambda$ acting by an $SO(N)$ generator, as it does not satisfy the double tracelessness condition, see footnote \ref{note:traceless}.
\end{appendices}
\bibliography{bound_on_a}

\end{document}